\documentclass[review,a4paper,fleqn]{cas-sc}

\usepackage[authoryear,round]{natbib}
\usepackage{multirow}
\usepackage{subcaption}
\usepackage{array}
\usepackage{gensymb}
\usepackage{amsmath}
\usepackage{lineno}
\usepackage{etoolbox}
\makeatletter
\patchcmd{\subsection}{\bfseries}{\itshape}{}{}
\patchcmd{\subsubsection}{\bfseries}{\itshape}{}{}
\makeatother

\definecolor{grey}{rgb}{0.5, 0.5, 0.5}

\usepackage{draftwatermark}

\SetWatermarkText{Preprint}

\SetWatermarkScale{1}
\SetWatermarkAngle{45}
\SetWatermarkColor{grey!10!white}

\def\tsc#1{\csdef{#1}{\textsc{\lowercase{#1}}\xspace}}
\tsc{WGM}
\tsc{QE}
\makeatletter
\renewcommand{\fnum@figure}{Fig. \thefigure}
\makeatother

\makeatletter
\renewcommand\section{\@startsection{section}{1}{\z@}%
  {-18pt plus -1ex minus -.2ex}
  {12pt plus 1pt}
  {\normalfont\Large\bfseries}} 
\renewcommand\subsection{\@startsection{subsection}{2}{\z@}%
  {-18pt plus -1ex minus -.2ex}
  {12pt plus 1pt}
  {\normalfont\itshape\large\bfseries}}
\renewcommand\subsubsection{\@startsection{subsubsection}{3}{\z@}%
  {-18pt plus -1ex minus -.2ex}%
  {12pt plus 1pt}%
  {\normalfont\itshape\normalsize\bfseries}}
\makeatother

\begin{document}
\let\WriteBookmarks\relax
\def\floatpagepagefraction{1}
\def\textpagefraction{.001}

\shorttitle{Future trends in Atlantic upwellings}    

\shortauthors{R. Flügel, S. Herbette et al.}  

\title[mode = title]{Spatial variation of future trends in Atlantic upwelling cells from two CMIP6 models}  



%

\author[1]{Raquel Flügel}[
    orcid=0009-0004-6614-2674
]

\cormark[1]


\ead{raquel.flugel@univ-brest.fr}



\affiliation[1]{organization={University of Brest},
            addressline={LOPS, IUEM, rue Dumont d'Urville}, 
            city={Plouzane},
            postcode={29280}, 
            country={France}}

\author[1]{Steven Herbette}

\author[1]{Anne-Marie Treguier}

\author[2]{Robin Waldman}
\affiliation[2]{organization={CNRM},
            addressline={42 Avenue Coriolis}, 
            city={Toulouse},
            postcode={31057}, 
            country={France}}

\author[3]{Malcolm Roberts}
\affiliation[3]{organization={Met Office},
            addressline={FitzRoy Road}, 
            city={Exeter},
            postcode={EX1 3PB}, 
            country={UK}}

\cortext[1]{Corresponding author. LOPS, IUEM, rue Dumont d'Urville, 29280 Plouzane, France}



\begin{abstract}
Eastern Boundary Upwelling Systems (EBUS) are characterized by wind-triggered upwelling of deep waters along the coast. They are hotspots of biological productivity and diversity and therefore have a high economic, ecological and social importance. In the past, different methods using surface data have been used to estimate upwelling. Recently, the IPCC has suggested directly assessing vertical velocities as a promising method. We use this method to study the two Atlantic EBUS from CMIP6 models from the HadGEM3-GC3.1 and the CNRM6-CM6 family, for both the historical period and a high-emission future scenario with spatial resolutions in the ocean component ranging from 1\degree to 1/12\degree. The two major upwelling regions are divided in subregions depending on their seasonality. The vertical transport index shows similar values to a wind-derived Ekman index. Directly evaluating upwelling from transport processes further provides information about the depth of the upwelling, which has previously been identified as an important factor for nutrient availability. We show that depending on the subregion of the upwelling system, different cell structures can be seen in terms of depth and distance to the coast of maximum velocities. When looking at possible future changes, high interannual variability limits the significance of the trends but could indicate a poleward shift of the upwelling regions. A detailed comparison of the spatial structures and the distinction in subregions is important to explain contradictory trends in previous works. 
\end{abstract}


\begin{highlights}
\item The Atlantic upwelling cells are diagnosed from the vertical velocity in the CNRM-CM6-1 and HadGEM3-GC3.1 coupled models.
\item Cell structures differ in terms of depth and distance to the coast depending on the subregion of each upwelling system.
\item Subregions also show different future trends, but few are significant.
\item The trends tend to support the hypothesis of a future poleward migration of upwelling systems.
\end{highlights}


\maketitle
\section{Introduction} \label{intro}
Eastern Boundary Upwelling Systems (EBUS) are found in the subtropical band, along the eastern margins of the major ocean basins. These coastal regions are characterized by dominant and semi-persistent equatorward trade winds blowing along-shore, and therefore inducing an offshore Ekman transport of the surface waters. The horizontal divergence of this transport, partly due to the no-normal-flow condition at the coast, drives the upwelling of cold, nutrient-enriched waters into the surface euphotic zone, which in turn, stimulates an intense primary production \citep{strub2013ocean}. Also contributing to the upwelling of deep waters, though at a lower rate, is the positive divergent oceanic circulation, which is created by cyclonic surface wind stress curl \citep{sylla2022}. These mechanisms contribute to making the four major EBUS of the World Ocean, {\it i.e} the Californian, the Humboldt, the Canary and the  Benguela upwelling regions, major productive ecosystems. While only covering 2 \% of the World Ocean's surface, these four systems account for $\sim$20 \% of global fish catch. Considering both the economic revenue generated by the fish catch and the impacts on the human food supply chain, recent studies estimate that $\sim$80 million people benefit directly or indirectly from today's functioning of the EBUS ecosystems \citep{GarciaReyes_UnderPressure,sylla2019weakening}.


Within the context of climate change, considering the future of EBUS therefore becomes of primary importance. Given that the major drivers of the upwelling in EBUS are the along-shore trade winds, the community has focused its efforts in considering projected wind changes.
Two main hypotheses are usually considered: i) the Bakun hypothesis \citep{bakun1990global} states that along-shore wind stress will increase in future climates because of the intensification of the inland low pressure systems on the eastern borders of these regions. Seawater heat capacity being higher than soil heat capacity, one expects the surface inland temperature to rise faster than the seawater one, and therefore the atmospheric thermal low over land to strengthen. ii) The second hypothesis states that there will be a poleward migration of the trade winds as the atmospheric Hadley cell expands poleward, driving in turn the poleward shift of the subtropical high pressure atmospheric cells found in the four ocean basins \citep{rykaczewski2015poleward}. 
While one expects a general increase of the volume of water that is being upwelled to the surface in the case of an overall increase of the trade winds (hypothesis i), along-shore regional disparities are expected to happen under the poleward migration hypothesis, with an intensification/decrease of the upwelling in the poleward/equatorward parts of the EBUS.

\cite{Gulev2021}'s recent analysis, based on historical datasets of the wind fields, led to a rejection of the Bakun hypothesis. Although the increase of surface temperature over the last century was higher over land (1.59 [1.34 to 1.83]°C) than over the ocean (0.88 [0.68 to 1.01]°C), this increase did not significantly impact the intensity of the upwelling favourable surface wind stress. On the other hand, the poleward migration of the Hadley cell was confirmed, and found to be more pronounced in the Northern than in the Southern Hemisphere. Additionally, contradicting trends were found for different EBUS. However, as already shown in the work of \cite{sydeman2014} and \cite{varela2015has}, trends appear to be very sensitive to the data product, period and geographical boundaries of the region of interest. This sensitivity highlights the lack of long-term robust observational datasets that can be used by the community to infer trends in upwelling systems.  

Further, the ocean circulation, forced in part by the wind stress curl, is known to play a major role in upwelling dynamics \citep{veitch2010modeling}. Thus, the future of EBUS does not depend on changes of the Ekman forcing only, but also on changes of the regional ocean circulation, which is geostrophic to first order. 

Future projections of EBUS rely on global coupled ocean-atmosphere circulation models \citep{IPCCRep}. 
The quality of these projections depends on the model's skill in reproducing the major dynamical features at play in the different EBUS, {\it i.e} a realistic along-shore wind stress and a dynamically consistent vertical upwelling cell. Models from the Coupled Model Intercomparison Project phase 6 (CMIP6, \cite{eyring2016}) display important biases of their Sea Surface Temperature (SST) in the upwelling systems, although higher resolution models perform better \citep{Varela2022}. Here we take advantage of the availability of high resolution simulations to go beyond previous studies that have assessed upwelling regions as a whole. Indeed, regional differences in upwelling intensity, seasonality and regional influencing factors can lead to local variations in current upwelling and future trends \citep{wang2015intensification}. 

\begin{figure*}[ht]
\centering
  \begin{subfigure}{0.5\textwidth}
    \centering
    \textbf{(a)}
    \includegraphics[width=0.75\textwidth]{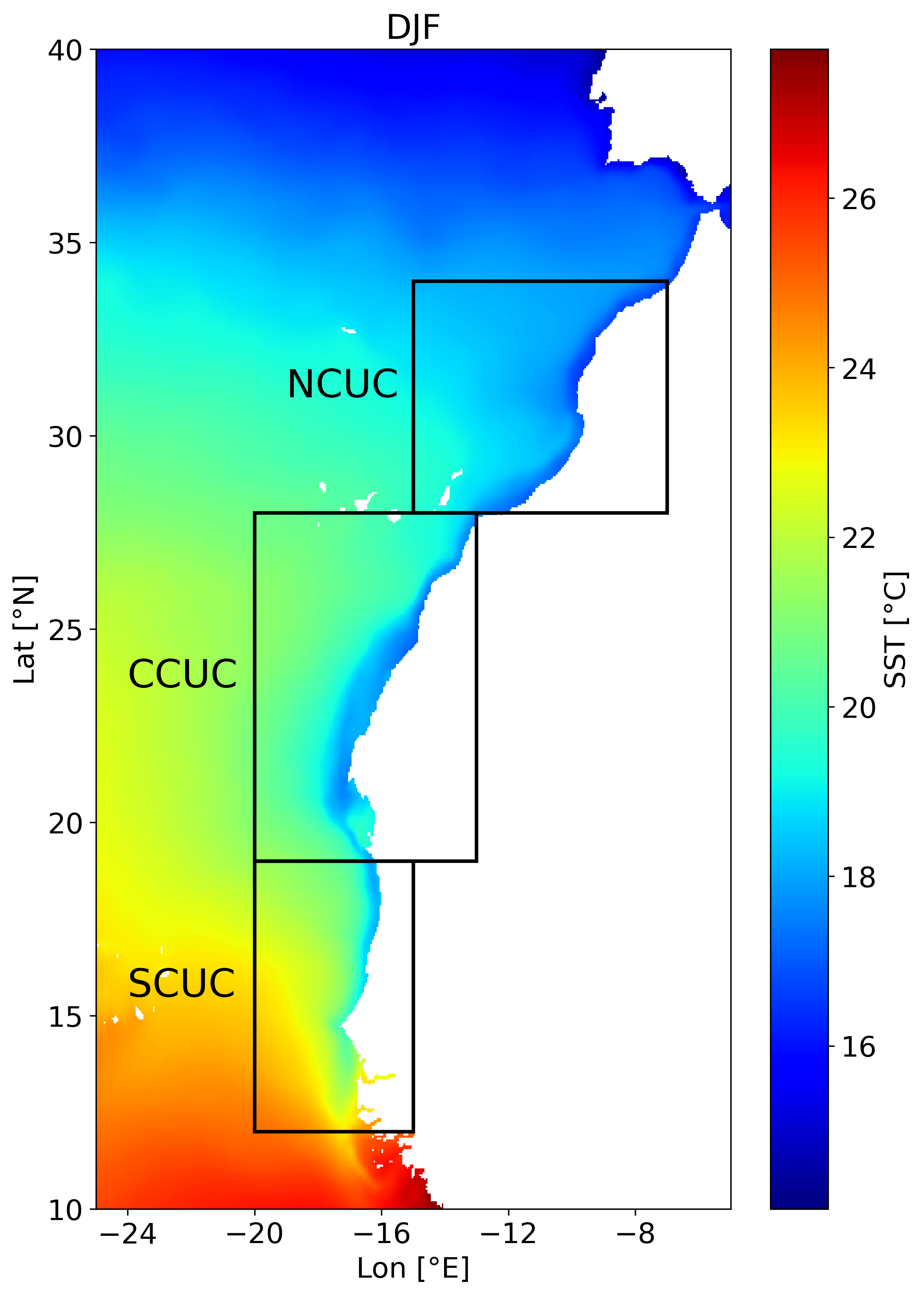}
    \label{subfig:sst_cci_a}
  \end{subfigure}%
  \begin{subfigure}{0.5\textwidth}
    \centering
    \textbf{(b)}
    \includegraphics[width=0.75\textwidth]{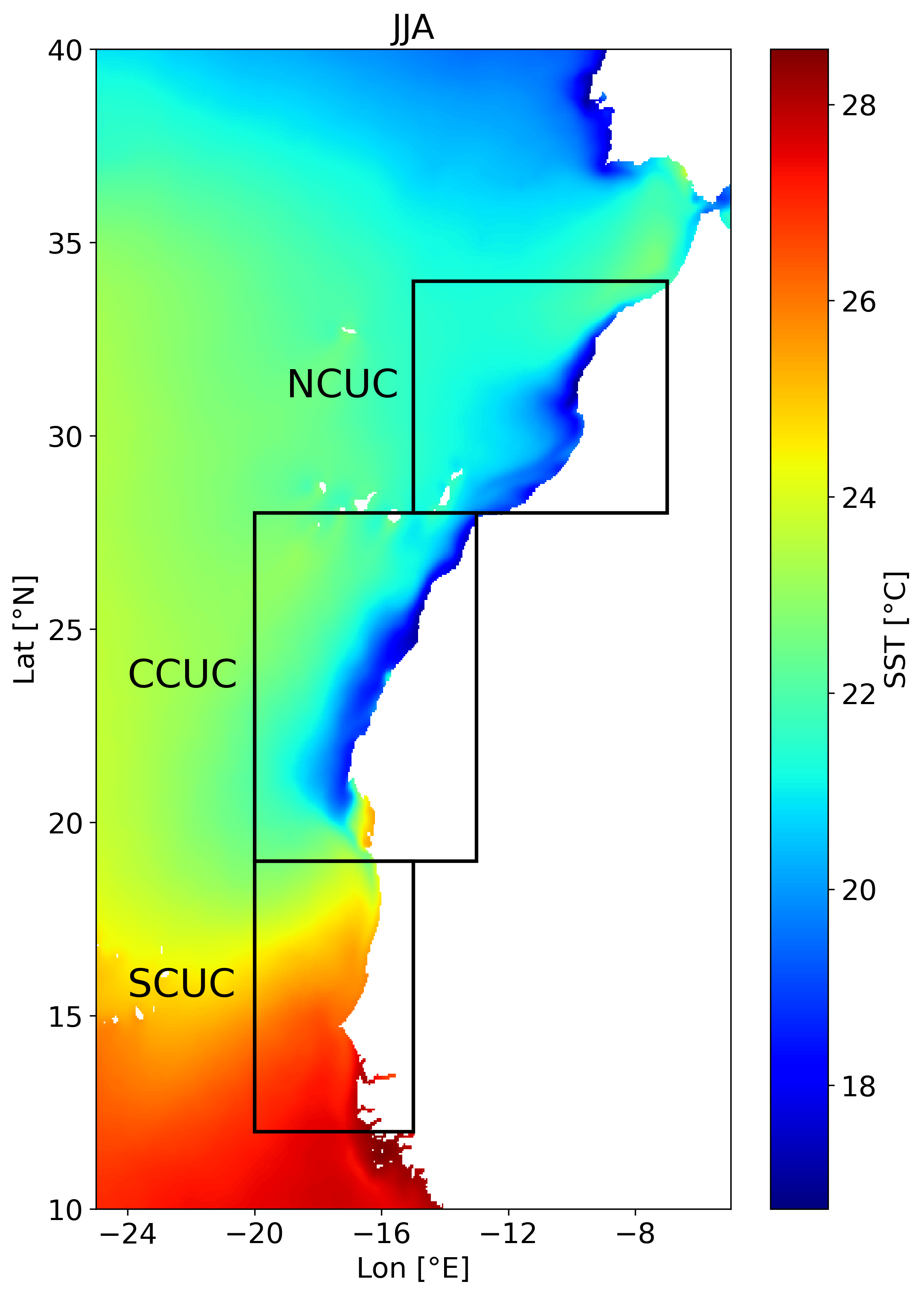}
    \label{subfig:sst_cci_b}
  \end{subfigure}
  
  \begin{subfigure}{0.5\textwidth}
    \centering
    \textbf{(c)}
    \includegraphics[width=0.75\textwidth]{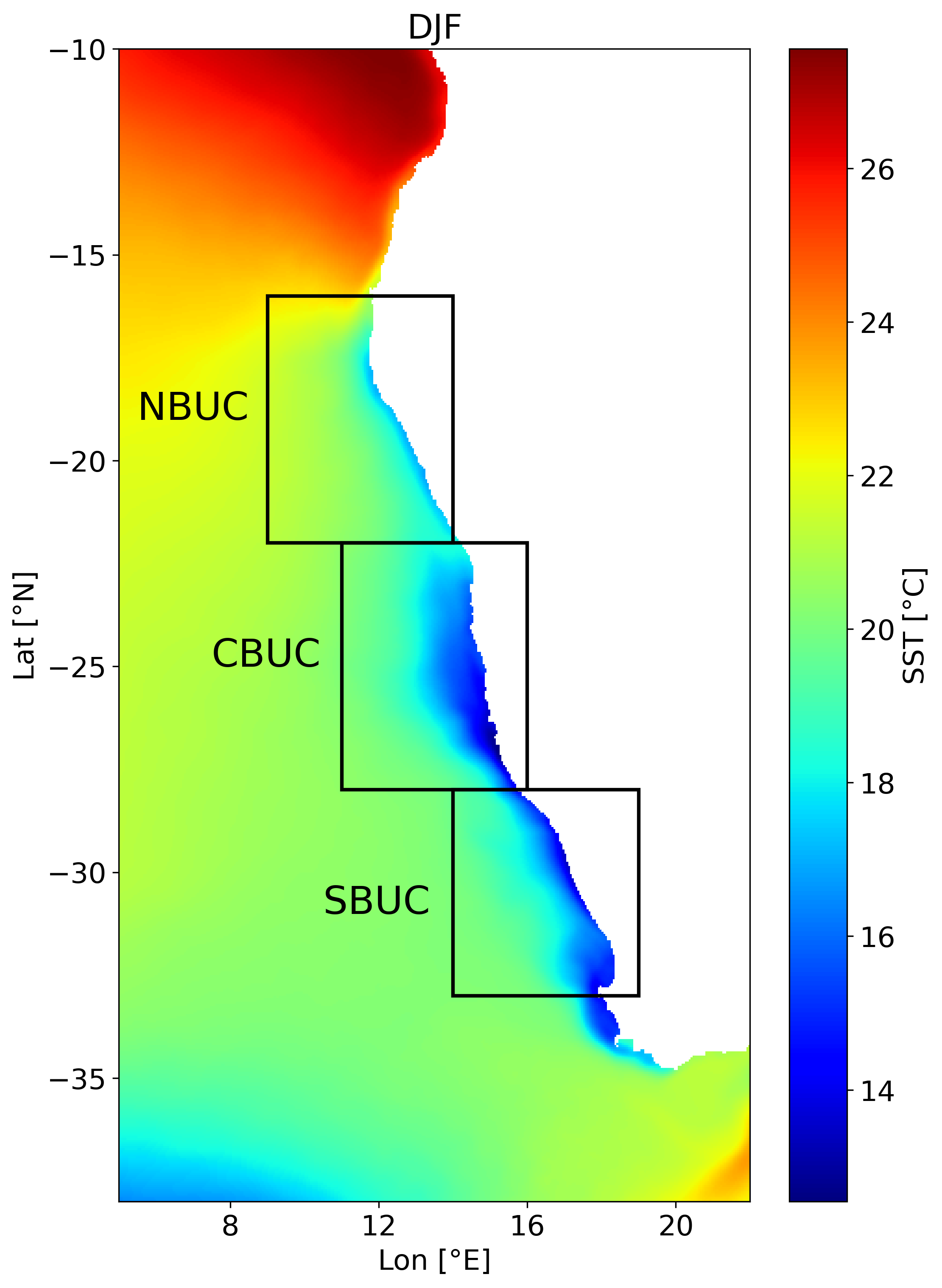}
    \label{subfig:sst_cci_c}
  \end{subfigure}%
  \begin{subfigure}{0.5\textwidth}
    \centering
    \textbf{(d)}
    \includegraphics[width=0.75\textwidth]{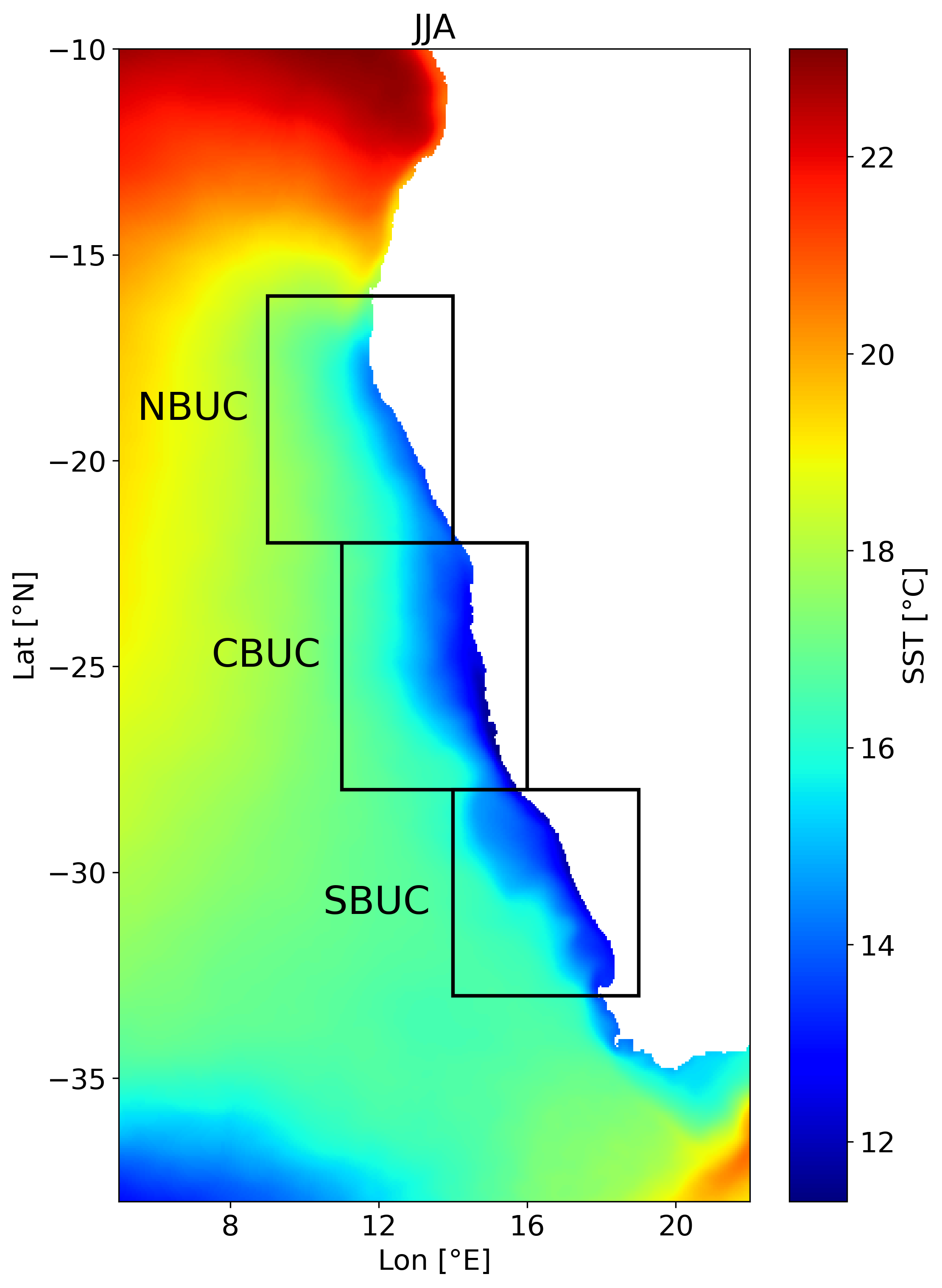} 
    \label{subfig:sst_cci_d}
  \end{subfigure}

  \caption{Seasonal mean Sea Surface Temperature from CCI Satellite data averaged over the historical 1995-2014 period for the Canary Upwelling region in boreal winter (a) and summer (b), and the Benguela Upwelling region in boreal winter (c) and summer (d). Black boxes show the subregions used for calculating indices.}
  \label{fig:sst_cci}
\end{figure*}

This work focuses on the Atlantic upwelling regions which can be seen in Fig. \ref{fig:sst_cci}. In the seasonal mean from European Space Agencies Climate Change Initiative (CCI) satellite SST observations, the colder SST upwelling signal along the coast reveals regional differences. 
The Canary region is located from 34°N to 12°N\ along the Northwest-African coast. The Northern Canary Upwelling Cell (NCUC) ranging from 28°N\  to 34°N\ is characterized by permanent weak upwelling with highest intensities in boreal summer and beginning of autumn. The Central Canary Upwelling Cell (CCUC), from 19°N\ to 25°N\, is a region of permanent strong upwelling with the highest upwelling rates in boreal summer. Figure \ref{fig:sst_cci} shows a larger contrast between coastal and offshore temperatures in the CCUC than in the NCUC, representing the different intensity of upwelling in the two subregions. The NCUC and CCUC summer upwelling is linked to the expansion of the Azores High and resulting wind strengthening \citep{aristegui2009sub} and is hence more pronounced in JJA (Fig. \ref{fig:sst_cci}b). In the Southern Canary Upwelling Cell (SCUC), from 12°N to 19°N, upwelling is seasonal and occurs only in boreal winter and spring (Fig. \ref{fig:sst_cci}a), when the Inter Tropical Convergence Zone (ITCZ) moves southward towards the equator \citep{aristegui2009sub, sylla2019weakening}. In boreal summer, the ITCZ moves northward and the resulting reversal in winds favors downwelling conditions. The cold SST signal is therefore not visible in the summer mean (Fig.\ref{fig:sst_cci}b).
The Benguela upwelling system is found along the Namibian and South African coast.  
The Northern Benguela Upwelling Cell (NBUC) experiences permanent upwelling with seasonal maxima in austral spring and winter (Fig. \ref{fig:sst_cci}d) and spreads over the region from 16°S to 22°S \citep{varela2015has}. The strongest upwelling region in the Benguela upwelling is the Central Benguela Upwelling Cell (CBUC). It includes the strong Lüderitz cell, and has the highest upwelling rates in austral summer \citep{HUTCHINGS200915}. In Fig. \ref{fig:sst_cci}, Panels c and d, this can be seen by a year-round cold SST signal with high zonal extent. This region is defined from 22°S to 28°S. In the Southern Benguela Upwelling Cell (SBUC), from 28°S to 33°S, upwelling occurs from austral autumn until spring, with a maximum in austral summer. While Fig. \ref{fig:sst_cci}, Panels c and d, both show a cold SST at the coast, the contrast between coastal and offshore temperature is higher in DJF, meaning stronger upwelling rates in this season. This separation in subregions is not only important for the distinction of the seasonal cycles and different impacting mechanisms, but different regions might also be impacted differently by the global climate change. \\
In the Canary upwelling cell, \cite{sylla2019weakening} used CMIP5 models to assess future changes in the Senegalo-Mauritanian upwelling region (12-20\degree N). They found a decrease of upwelling revealed by both SST and Ekman based indices in the southern region and unsignificant changes (Ekman index) or a decrease of upwelling (SST index) in the northern region. 
Recently, \cite{jing2023geostrophic} and \cite{chang2023uncertain} evaluated future trends in upwelling regions using coupled global simulations performed with a high resolution coupled climate model (1/10° ocean resolution). Although these two studies use the same simulations, their conclusions for future trends in the Benguela region differ. While \cite{chang2023uncertain} find negative trends for vertical velocities and positive changes in Ekman transport, \cite{jing2023geostrophic} find an increase in all indices.

These results highlight the necessity to consider more models for studying the oceans response to the external forcing. 
We will also show that partitioning the upwellings into subregions helps interpret the apparent contradictions between \cite{jing2023geostrophic} and  \cite{chang2023uncertain}. Our analysis is based on vertical transport because it is a direct measure of total upwelling. It includes the contribution of all physical processes creating upwelling. We introduce a new representation of the integrated transport that provides information about the characteristics of upwelling cells: strength, depth and distance to the coast. 

In section 2, we explain the methods and present the model and observational data used in this work. In section 3, we compute the seasonal cycle of upwelling indices during the historical period and compare them to observations in order to validate the models. The structure of the upwelling cells is analyzed in relation with the wind and SST indices, and future trends within the highest emission scenario are presented and discussed. The conclusion is drawn in section~4. \\
\newline

\section{Data and methods}

\subsection{CMIP6 model data}
The following analysis uses  monthly outputs of five simulations from the Coupled Model Intercomparison Project phase 6 (CMIP6). Three simulations come from the UK Met Office's HadGEM3-GC3.1 model \citep{Andrews2020}, and two from the Centre National de Recherches Météorologiques (CNRM) CNRM-CM6-1 models \citep{Voldoire2019c}. For consistency, as our analysis focuses mainly on the ocean response to upwelling favourable winds, all the five above-mentioned simulations were run using the same oceanic component, {\it i.e.} NEMO3.6 with 75 vertical levels. Three different resolutions are considered in the simulations. A low ocean resolution ($\sim$70~km, 1°) and a medium ocean resolution ($\sim$20 km, 1/4°) run were analysed for both the UK Met office and CNRM model over the  historical (1995-2014) and future (2015-2100) period under SSP5-8.5 \citep{eyring2016, oneill2016}. For the highest ocean resolution model, HadGEM3-GC3.1-HH ($\sim$8 km, 1/12°), run in the context of HighResMIP \citep{haarsma2016}, the future period is only available until 2050  (highres-future simulation under scenario RCP8.5). 
The limited number of simulations considered is purely linked to the availability at the time of analysis, in the ESGF online archive, of monthly outputs for meridional wind stress $\tau_y$, potential temperature $\Theta$ and vertical water mass transport $wmo$  \citep{griffies_2016}. Note that we only use one member per model for comparability, as for some models, only one realization is available. 
\newline

\subsection{Observation data}
The performance of the CMIP6 models is first evaluated over the historical period by comparing their meridional wind stress and SST to the ones provided by the European Centre for Medium-Range Weather Forecasts (ECMWF) ERA5 Reanalysis and the European Space Agencies Climate Change Initiative (CCI), respectively. ERA5 is a global atmospheric reanalysis with a 0.25\degree\ spatial resolution using 4D-Var data assimilation and model forecasts and is available over the time period 1940~- present \citep[{[dataset]}][]{Hersbach2018,CopernicusC3S2023}. CCI is a level-4 satellite product, with a spatial resolution of 0.05\degree, covering 35 years over the 09/1981~- 12/2016 time period. The CCI SST \citep{Merchant2019} uses both Advanced Very High Resolution Radiometer (AVHRR) and Along Track Scanning Radiometer (ATSR) Sea Surface temperature records, \citep[{[dataset]}][]{Good2019}.
\newline

\subsection{Indices} \label{sec:indices}
\subsubsection{SST and Ekman Upwelling Indices}
The SST (SI) and Ekman (EI) Upwelling Indices are two classical indices commonly used in the literature to characterize the strength of the wind-driven upwelling and the ocean response over time \citep{varela2015has,sylla2019weakening,rykaczewski2015poleward}. They are computed as follows:
\begin{eqnarray}
\centering
    SI(y) & = SST_{\mbox{offshore}}(y) - SST_{\mbox{coast}}(y),  \label{eq:sst_ind} \\
    EI(y) & = \frac{\tau_y(y)}{\rho_0 f(y)}, \label{eq:ekman}
\end{eqnarray}
where $SST_{\mbox{offshore}}$ and $SST_{\mbox{coast}}$ are latitude dependant ($y$) sea surface temperatures taken offshore and inshore of the coastal upwelling band, respectively. $\tau_y$ is the meridional surface wind stress, $f$ the Coriolis parameter, and $\rho_0 = 1025 $~kg~m$^{-3}$ a constant reference density. In theory, positive $SI$ should be directly linked to the cooler SSTs found near the coast in the event of a cold deep-water upwelling. In practice, $SST_{\mbox{coast}}$ cannot always be taken closest to the coast, as SST in shallow waters is mainly driven by surface heat fluxes, and therefore does not necessarily reflect the occurrence of upwelling.  Following the work of \cite{BENAZZOUZ201438} in the Canary Upwelling System, the coastal reference point is taken on the 100~m isobath. $SST_{\mbox{offshore}}$ is subsequently taken $5$\degree ($\sim$500 km) offshore \citep{CROPPER2014}. The $EI$, first defined by \cite{bakun1973coastal}, corresponds to the net offshore zonal Ekman transport. As coastline orientation is approximately north-south in most EBUS, it provides a good assessment of the total cross-shore Ekman transport \citep{rykaczewski2015poleward}. $\tau_y$ is taken at 1\degree~offshore in all models for consistency. This distance is large enough to capture the maximum along-shore wind speeds, and therefore may also be interpreted as the offshore limit of the coastal band over which wind-driven coastal upwelling occurs.
$SI$ and $EI$ are subsequently averaged in latitude in order to provide single values indices for each subregion of the the Canary and Benguela upwelling systems defined in Fig.~\ref{fig:sst_cci}.

\subsubsection{Vertical transport and vertical transport index}
\begin{figure}[ht]
\begin{center}
  \begin{subfigure}{0.5\textwidth}
    \textbf{(a)}
    \end{subfigure}
    \begin{subfigure}{0.5\textwidth}
    \centering
    \includegraphics[width=0.75\textwidth]{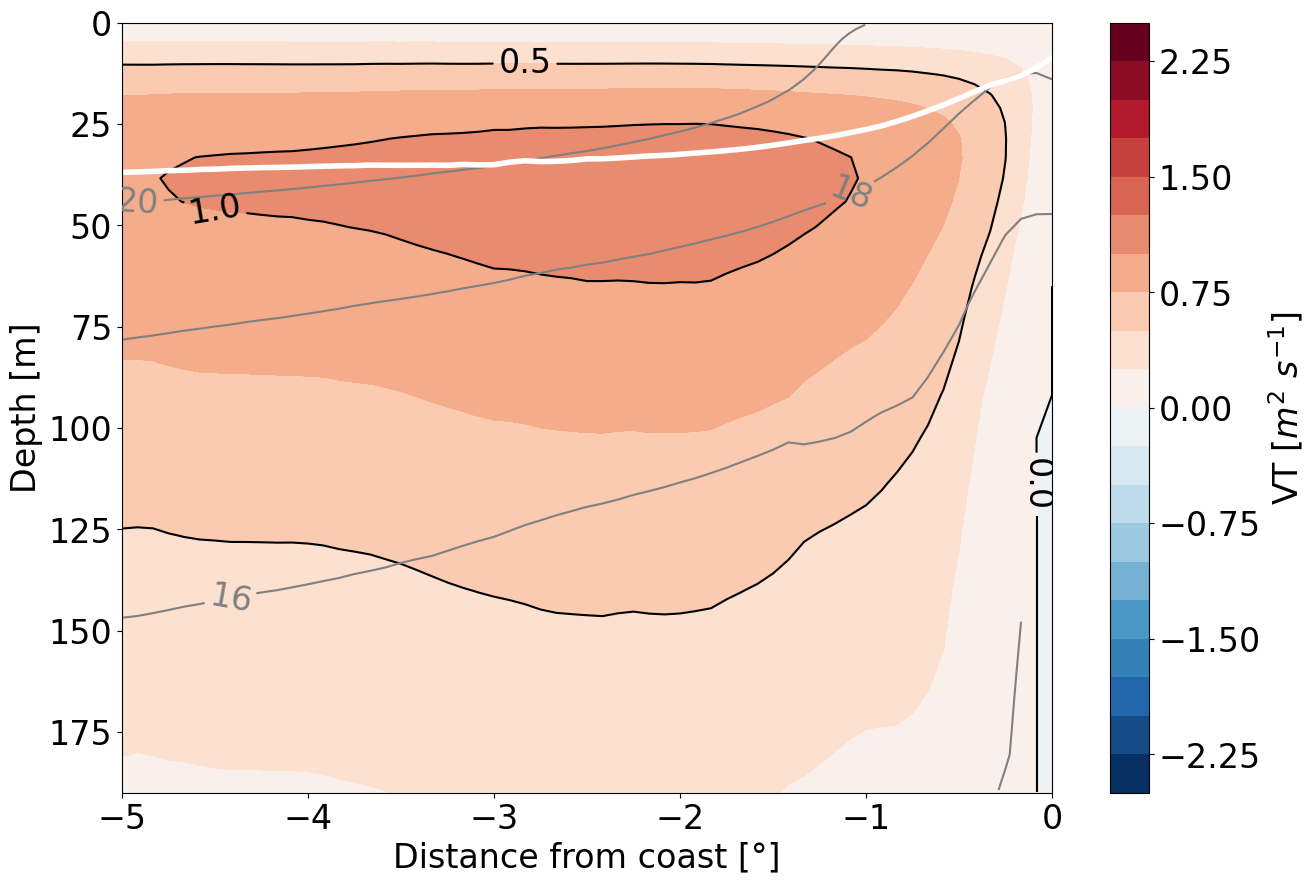}
    \label{subfig:cell_annual_a}
  \end{subfigure}
  \begin{subfigure}{0.5\textwidth}
    \textbf{(b)}
    \end{subfigure}
    \begin{subfigure}{0.5\textwidth}
    \centering
    \includegraphics[width=0.75\textwidth]{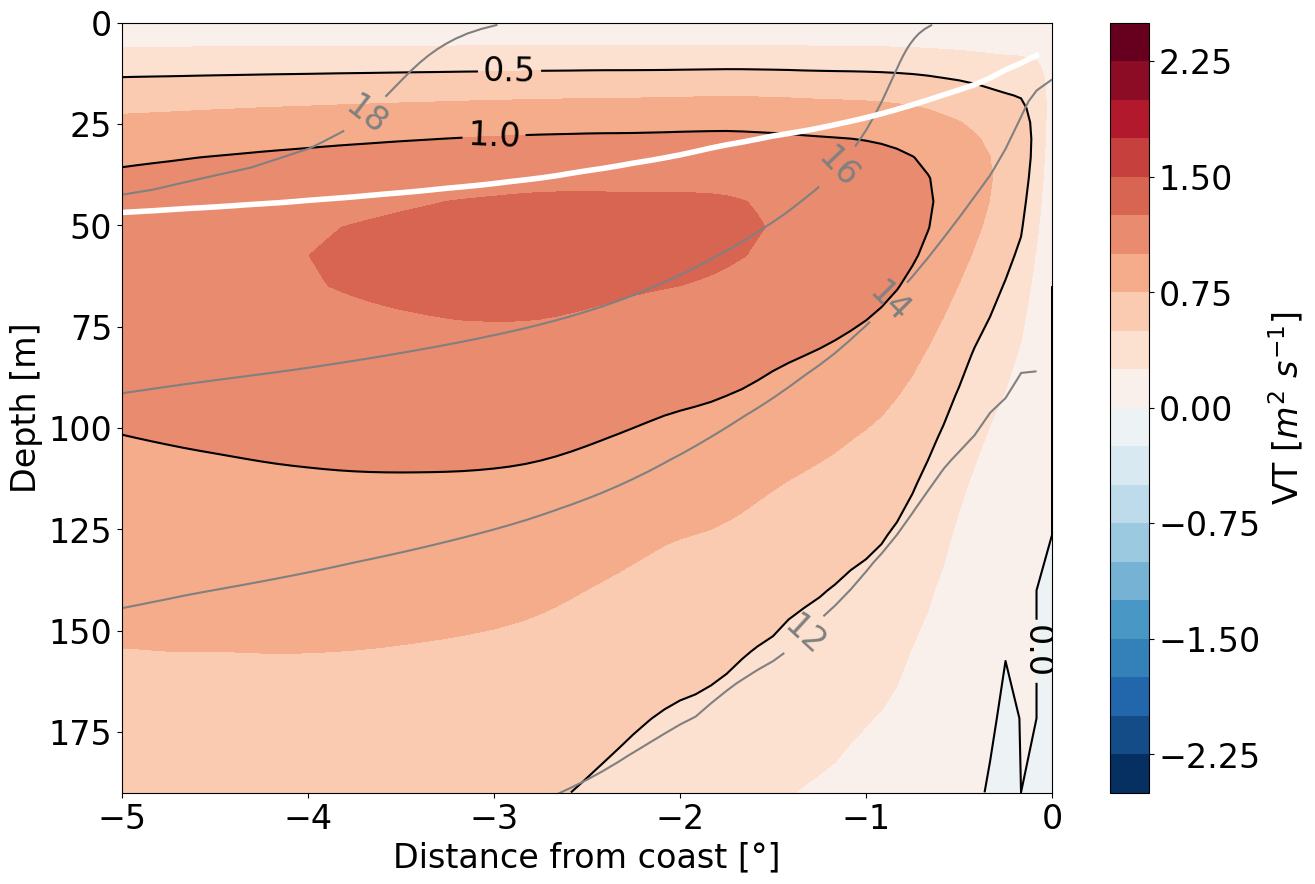}
    \label{subfig:cell_annual_b}
  \end{subfigure}
  \caption{Annual mean vertical transport $VT$ (color-filled black contours in m$^{2}$~s$^{-1}$) averaged in time over the historical 1995-2014 period and in latitude over the entire region retained for the:  (a) Canary Upwelling; (b) Benguela Upwelling. Contours of potential temperature (solid grey in $^\circ$C) and mixed layer depth (solid white) are superimposed. The model shown here is HadGEM3-GC3.1-HH.}
  \label{fig:cell_annual}
\end{center}
\end{figure}

The Vertical Transport ($VT$) computes the total volume of water being upwelled ($VT>0$) or downwelled ($VT<0$), per unit of meridional distance, at depth $z$, within a coastal band of width $X$. It is calculated (Eq.~\ref{eq:wmo}) by summing $wmo$ of the CMIP6 archive from the coast to an offshore distance $X$ in the zonal direction and dividing by the meridional size of the cell $\delta y$ and the reference sea-water density $\rho_0$.
\begin{equation}\label{eq:wmo}
\centering
VT(X,y,z) = \frac{1}{\rho_0}\sum_{i_{coast}}^{i_{X}} \frac{wmo(x_{i},y,z)}{\delta y}
\end{equation}
$VT$ is expressed in the same unit as the Ekman transport (m$^2$~s$^{-1}$). Within an idealized 2D depth-cross-shore distance model, $VT$ would simply be interpreted as the streamfunction of a flow that leaves the highest iso-$VT$ contours on its right. Although the realistic CMIP6 simulations do not strictly satisfy the condition of zero along-shore divergence, $VT$ still provides a meaningful visualization of the strength and vertical structure of the upwelling cells, especially when it is averaged seasonally and spatially along-shore over the different sub-regions. The annual and along-shore averages of $VT$ are plotted in Fig.~\ref{fig:cell_annual} for the Canary and Benguela upwelling systems. The upwelling cells are clearly represented. Both regions show the maximum $VT$ to be located right below the mean Mixed Layer Depth (MLD). However, the Canary upwelling cell is less intense, more surface intensified and closer to the shore than its counter part in the southern hemisphere. Although this figure only shows $VT$ for the HadGEM3-GC3.1-HH model, the four other models have a comparable vertical structure, despite some differences in the intensity, vertical extension, and offshore position of the cells.


$VT$ is subsequently used to construct one other single value upwelling index that may be compared to $SI$ and $EI$, but that better represents the volume of waters being uplifted towards the surface within a coastal band. For consistency with $EI$, this coastal band is considered to extend as far as $1^\circ$ offshore $X_{1^\circ}$ \citep{tim2015decadal}. The Vertical Transport Index ($VTI$) is additionally defined as the vertical transport $VT$ at a fixed depth of 50~m \citep{jing2023geostrophic}:
\begin{equation}
\begin{array}{rlc}
VTI(y) & = & VT(X_{1^\circ},z_{50~\mbox{m}}) \\
VT_{m}(y) & = & VT(X_{1^\circ},z_{\mbox{max VT}})
\end{array}
\end{equation}
Note that $VTI$ is also averaged along-shore over each sub-region, as $SI$ and $EI$. When looking at the cell structure in section 3.3, the transport $VT_{m}$ is taken at the depth where $VT$ reaches its maximum.
\newline

\section{Results and discussion}
\subsection{Influence of model resolution on the SST bias}\label{sstbias}
\begin{figure*}[ht]
  \begin{subfigure}{0.49\textwidth}
    \centering
    \includegraphics[width=\textwidth]{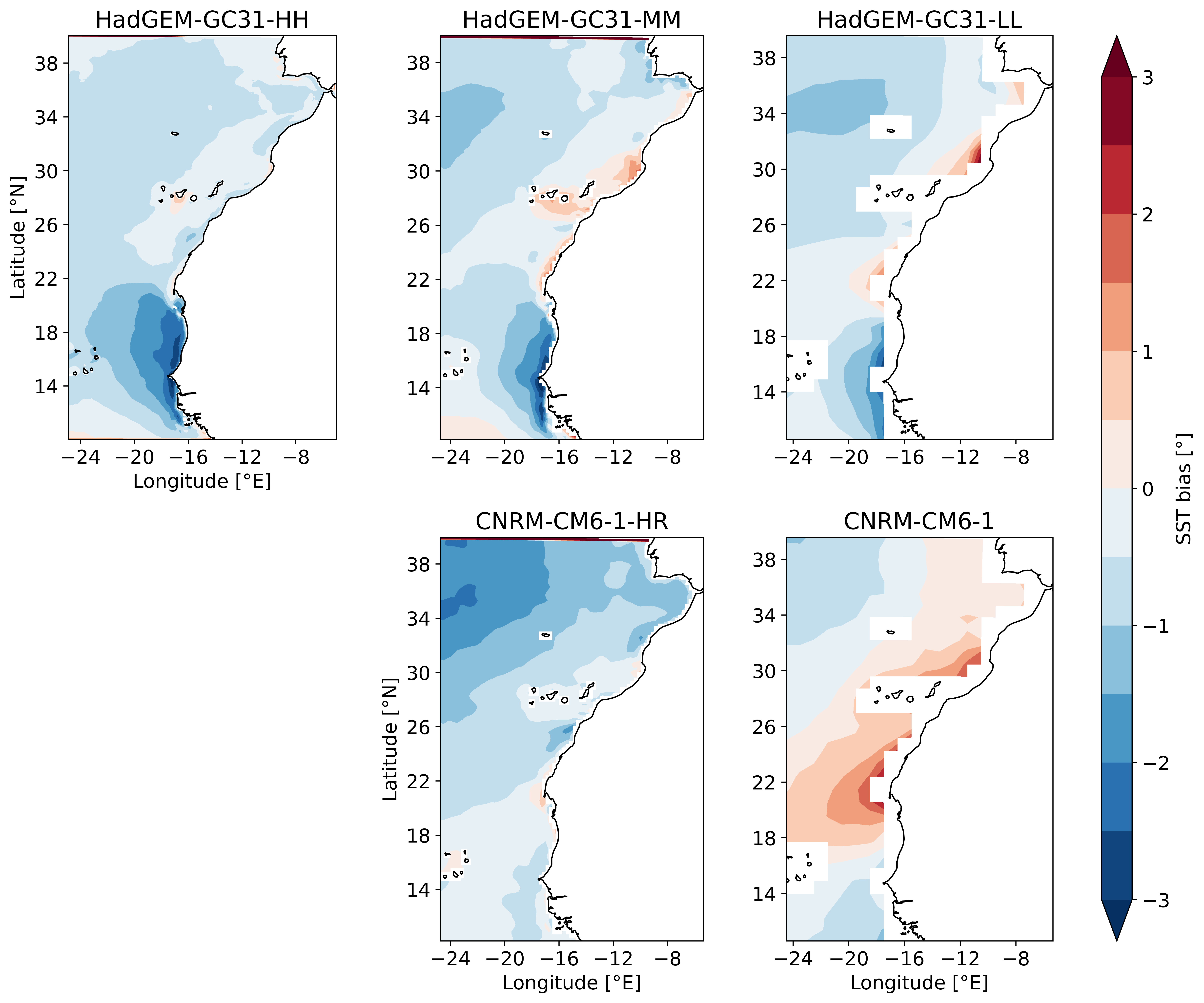}
    \caption{}
    \label{subfig:sst_bias_a}
  \end{subfigure}
  \begin{subfigure}{0.49\textwidth}
    \centering
    \includegraphics[width=\textwidth]{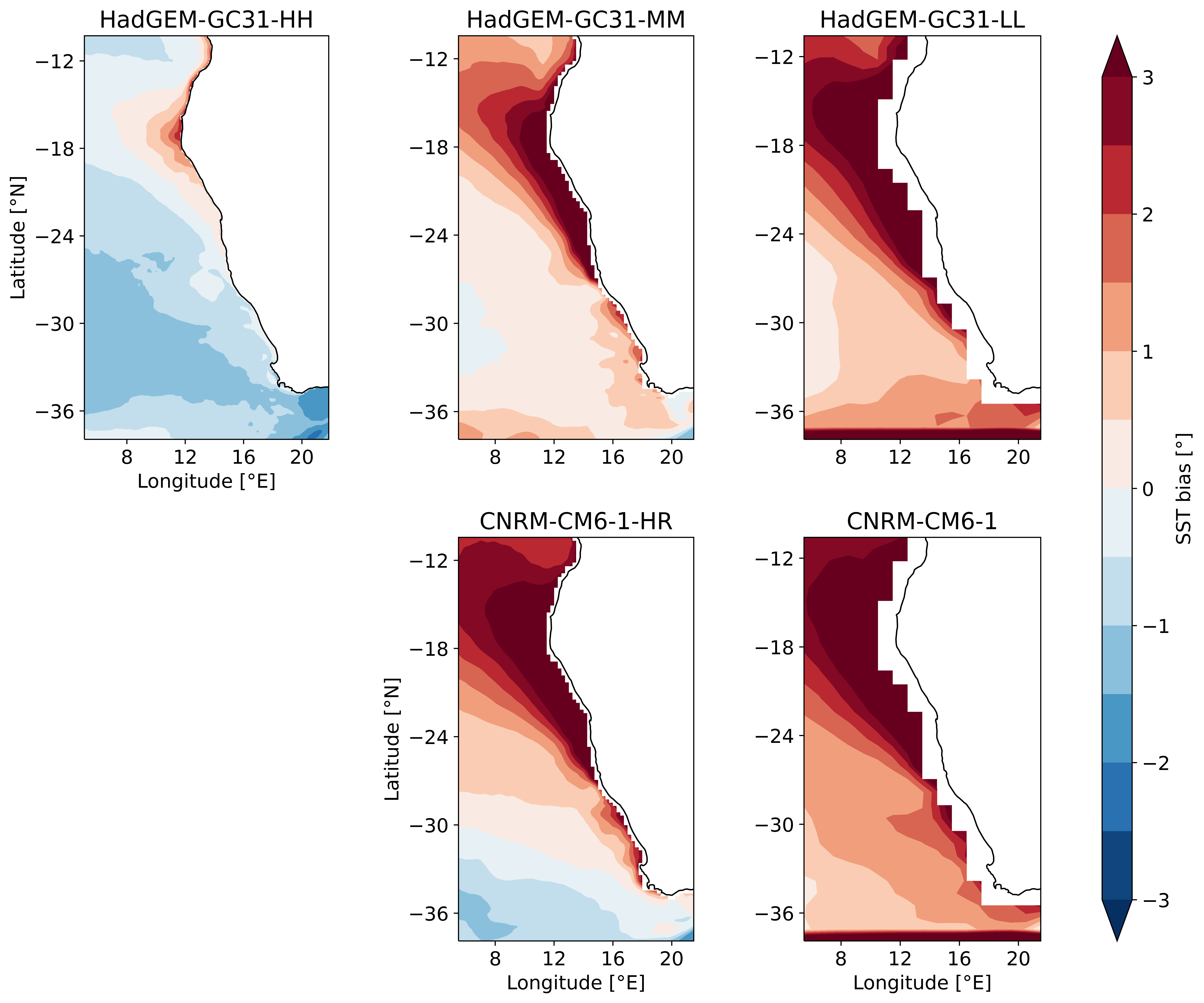}
    \caption{}
    \label{subfig:sst_bias_b}
  \end{subfigure}
  \caption{Annual mean SST bias showed by the CMIP6 models over the 1995-2014 period in the Atlantic EBUS: (a) Canary; (b) Benguela. $SST_{\mbox{bias}} = SST_{\mbox{CMIP6}}- SST_{\mbox{CCI}}$.}
  \label{fig:sst_bias}
\end{figure*}

The behavior of the CMIP6 models in the two EBUS of the Atlantic is first assessed by comparing their SST to the satellite CCI SST. Like most Global Coupled Models (GCMs) \citep{strub2013ocean}, all five CMIP6 models tend to overestimate the SST within the coastal band of the Canary and Benguela (Fig.~\ref{fig:sst_bias}).

For all models except HadGEM3-GC3.1-HH, SST biases are found to be much higher in the Benguela region than in the Canary. Within the Canary system, biases do not exceed +2$^\circ$C in the subtropics (lat>20$^\circ$N), and may even become sightly negative in the tropics. Within the Benguela system, biases exceed 3$^\circ$C in the subtropics, and gradually increase towards the equator until reaching 9$^\circ$C in the tropical southern Atlantic. Although one would expect models with a higher resolution to perform better in representing the SST \citep{park2020resolution}, increasing the model's resolution did not systematically improve the model's SST in the 5 simulations considered here. Within the Canary system, as resolution gets higher, the warm subtropical bias is reduced in the subtropics, but the cold bias can get worse in the tropics. Within the Benguela, the warm bias remains as high in the 20~km resolution model as in the 70~km models at all latitudes. Nevertheless, the HadGEM3-GC3.1-HH shows an above average performance in regards to the southern Benguela region, with a maximum positive bias of 3\degree C in the northern region and a slight underestimation of SST of up to 1\degree C in the southern region. This may be seen as  a remarkable performance as such bias is lower than the one of the 1/10\degree resolution Community Earth System Model High resolution model (CESM-H) \citep{jing2023geostrophic,chang2023uncertain}. \cite{richter2020overview} already found that the HadGEM family to be one of the best performing in the reduction of the Tropical Atlantic SST error. 

These results confirm that the SST biases of the 5 simulations here considered are in the range of values commonly found in previous studies. A misrepresentation of the SST in the southern Canary upwelling with erroneous year-round cold SSTs at the coast has been observed in several CMIP5 models in \cite{sylla2019weakening}, which they ascribe to a misrepresentation of the position Intertropical Convergence Zone. Warm biases, commonly found in the coastal band of subtropical EBUS, are generally attributed to erroneous radiation fluxes and lower than observed upwelling favourable wind intensity, the two being possibly linked to a misrepresentation of the cloud cover \citep{richter2020overview}.
Although they may reach sometimes very high values, these biases are believed not to cause major losses in the prediction skills of GCMs, as long as models can capture the existing variability \citep{park2020resolution,richter2020overview}. Further, when it comes to long-term future predictions in upwelling regions, GCMs are the only available option \citep{sylla2019weakening}. 
\newline

\subsection{Seasonal cycle} \label{seas}
\begin{figure*}[ht]
    \includegraphics[width=\textwidth]{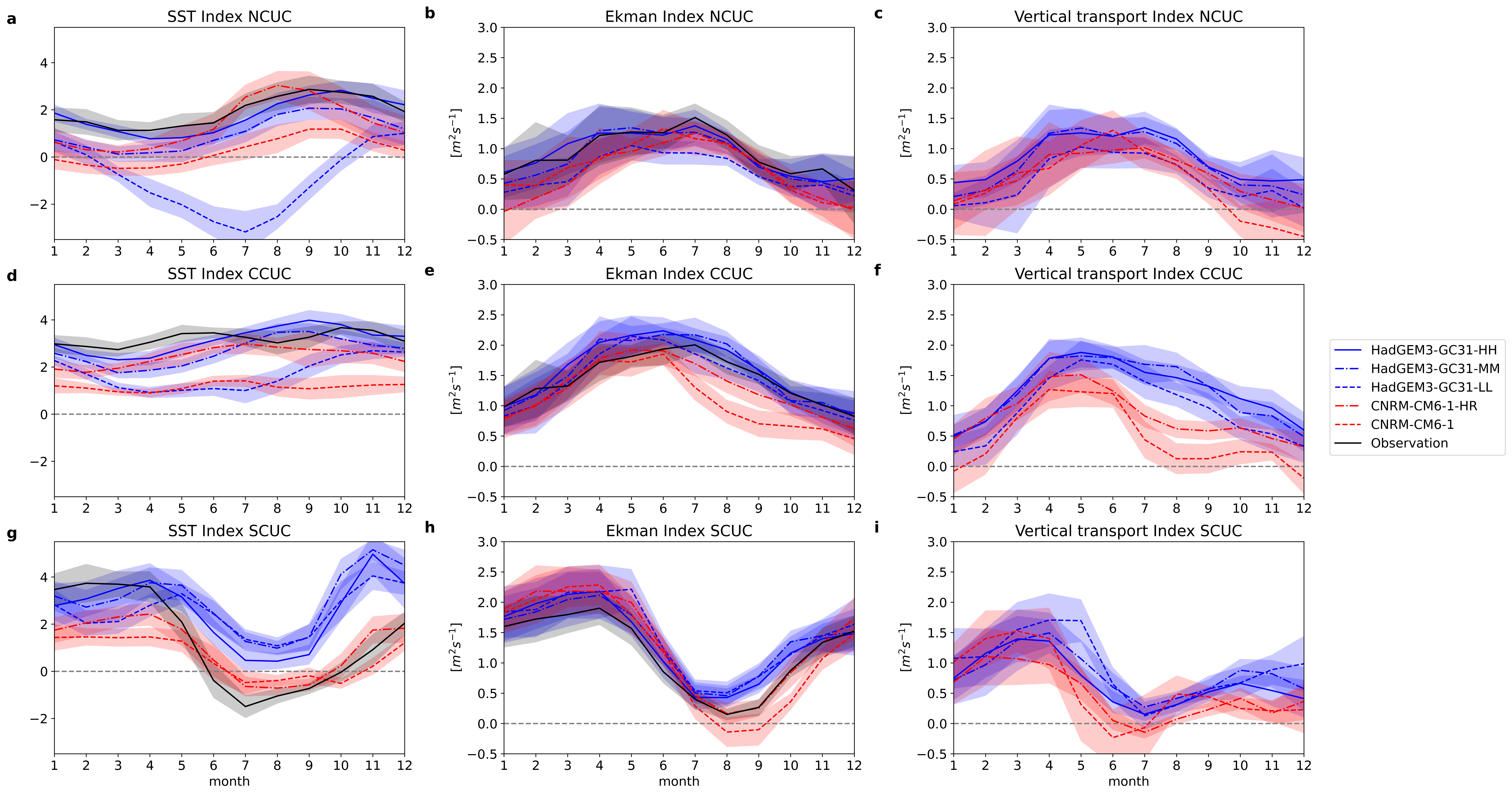}
  \caption{ Mean monthly upwelling indices over the time period 1995-2014 and their standard deviation for the three subregions in the Canary Upwelling system for the five CMIP6 models, the SST observations from CCI (Panels a,d,g) and the ERA5 Reanalysis wind stress (Panels b,e,h).}
  \label{fig:seasonal_cycle_can}
\end{figure*}

\begin{figure*}[ht]
    \includegraphics[width=\textwidth]{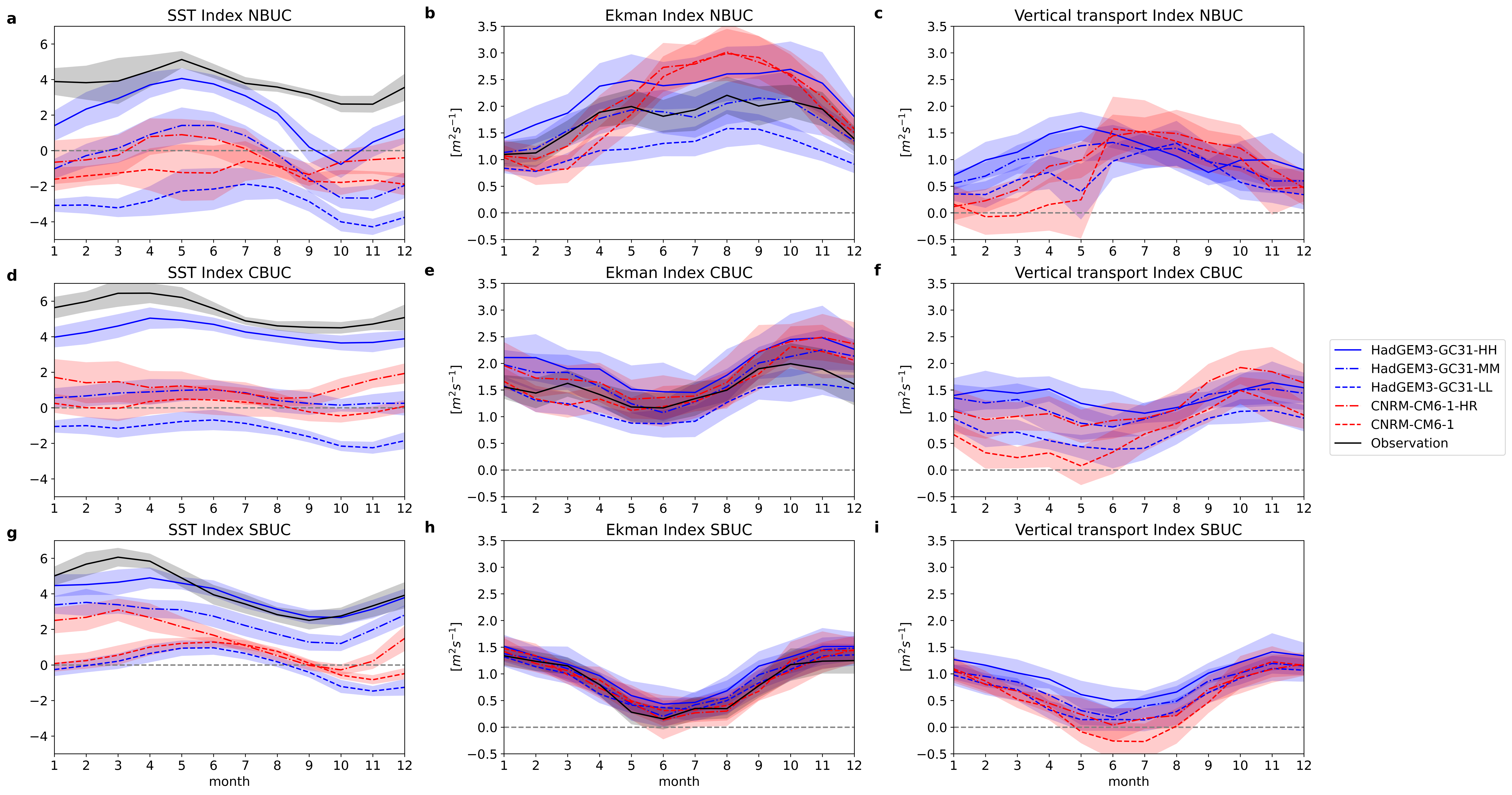}
  \caption{Mean monthly upwelling indices over the time period 1995-2014 and their standard deviation for the three subregions in the Benguela Upwelling system for the five CMIP6 models, the SST observations from CCI (Panels a,d,g) and the ERA5 Reanalysis wind stress (Panels b,e,h).}
  \label{fig:seasonal_cycle_ben}
\end{figure*}

In the past, it has been found that models do not always capture the seasonal upwelling onset correctly \citep{varela2015has, richter2020overview, sylla2019weakening}. In the Atlantic, the southernmost regions of the Canary and Benguela upwellings have specific seasonal dynamics \citep{strub2013ocean, sylla2019weakening}. We therefore assess the climatologies of the seasonal cycle using the subregions defined in Fig. \ref{fig:sst_cci}. We compare the models to observational data for the surface upwelling indices which helps to further evaluate the model's performance. 
We then introduce the $VTI$ derived from vertical water mass transport and compare it to the surface indices. The seasonal cycle for the three indices and for each of the subregions can be seen in Fig. \ref{fig:seasonal_cycle_can} for the Canary Upwelling system and in Fig. \ref{fig:seasonal_cycle_ben} for the Benguela Upwelling system. 
The seasonal cycle of the SST index from the models and CCI satellite observation are shown in Fig. \ref{fig:seasonal_cycle_can}, Panels a,d,g. In the NCUC HadGEM3-GC3.1-LL shows a negative SST index in boreal summer (Fig. \ref{fig:seasonal_cycle_can}a). This is due to strong coastal warming along the Moroccan coast in this model during the summer months. The reasons for this have not been explored further. Apart from this model, the models display the onset of upwelling in early summer in the NCUC compared to CCI, although underestimating its intensity. In the CCUC (Fig. \ref{fig:seasonal_cycle_can}d) all models portray the year round positive SST Index also found in CCI, but fail to represent its intensity and its maxima. However, in Fig. \ref{fig:seasonal_cycle_can}g we see that upwelling onset is portayed rather well, where in previous work using the SST index from CMIP5 models in the SCUC, the models (notably the previous generation of the two models used here) have failed to reproduce the seasonal cycle and the onset of upwelling \citep{sylla2019weakening}. 
In the Benguela Upwelling System, the SST Index shows negative values for several models, especially in the NBUC (Fig. \ref{fig:seasonal_cycle_ben}, Panels a,d,g). This is linked to the Tropical Atlantic SST bias (see section \ref{sstbias}, Fig. \ref{fig:sst_bias}), which is more intense at the coast. As expected from the analysis of the SST bias, higher resolution models also perform better in depicting the SST index than the lower resolution members of the same family, with HadGEM3-GC3.1-HH being the closest to observations. Upwelling onset is delayed compared to CCI, with the delay becoming more pronounced as the the index becomes lower. 

While for the SST index the models show a large difference in terms of amplitude and upwelling onset, the Ekman index (Figs. \ref{fig:seasonal_cycle_can} and \ref{fig:seasonal_cycle_ben}, Panels b, e, h) is closer to the observations in both upwelling systems. This has already been found in previous work in the SCUC \citep{sylla2019weakening}. The Ekman Index follows the same seasonal cycle and the differences between the models' performance are small. Nonetheless, the Ekman transport is overestimated in the NBUC in austral winter by most models compared to observations. Previous work has shown that an improvement of model SST in this region often goes along with overestimation of wind \citep{richter2020overview}. This is consistent with the overestimation in HadGEM3-GC3.1-HH winds in relation with its comparably good performance in the SST index in the Benguela upwelling system. However, this does not explain the overestimation in the two CNRM models. HadGEM3-GC3.1-MM shows a very good performance in all three subregions in the Benguela for the $EI$. When comparing the Ekman index with the SST index an obvious difference between the seasonal cycle and the onset of upwelling becomes evident. This is the case for modeled as much as the observed data. Upwelling onset is observed later in the season for the SST index than for the other indices. A later onset of upwelling from a SST derived compared to an Ekman divergence derived index has also been observed by \cite{sylla2019weakening}. \\

When comparing SST and Ekman indices to the Vertical Transport index (Panels c,f,i), the seasonal cycle from $VTI$ appears very close in intensity and phase to the $EI$. The $EI$ can be compared directly to $VTI$, as they have the same units. The seasonal cycle of the $VTI$ behaves very similar to the $EI$ for both upwelling regions and all subregions. This shows that there is a direct oceanic response to the wind forcing, which is reproduced by the models dynamics. Apart from a few extrema in the Benguela Upwelling Region (Fig. \ref{fig:seasonal_cycle_ben}), the CNRM models tend to show lower vertical transport rates than the HadGEM models. This is in accordance with the $EI$. The difference in intensity between $EI$ and $VTI$ is highest the closer to the equator in both regions (Fig. \ref{fig:seasonal_cycle_can}i and Fig. \ref{fig:seasonal_cycle_ben}c), while in the poleward portions (Fig. \ref{fig:seasonal_cycle_can}c and Fig. \ref{fig:seasonal_cycle_ben}i) intensities and seasonal cycle of $EI$ and $VTI$ correspond almost perfectly. This indicates that other transport processes might be more important in the equatorward portions of the upwellings. This could be linked to the convergence of surface flows within the coastal band, onshore or along-shore, due to the large-scale circulation \citep{veitch2010modeling}. 
\newline

\subsection{Vertical structure of the upwelling cell}

\begin{figure*}[ht]
    \includegraphics[width=\textwidth]{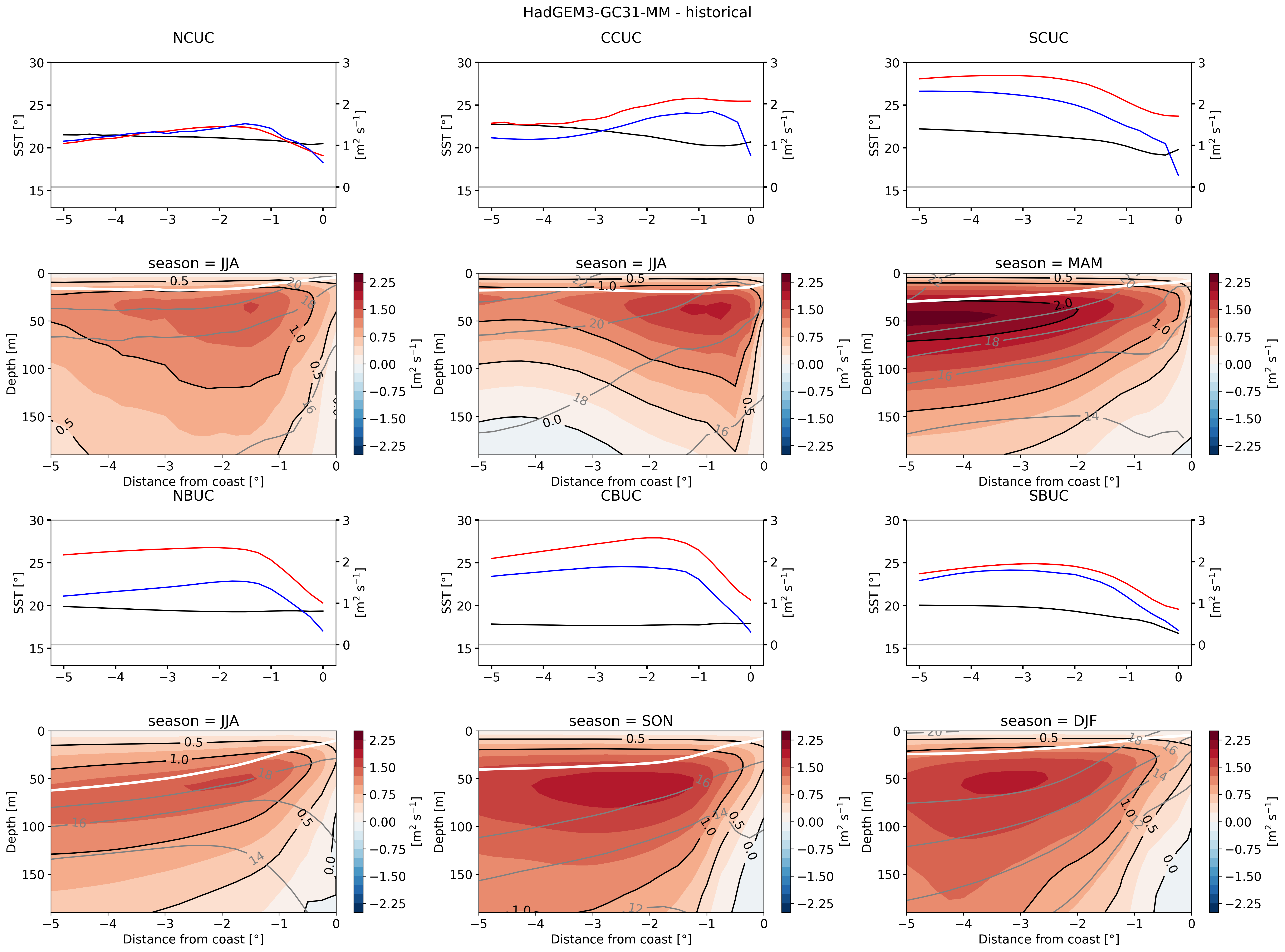}
  \caption{Cross-shore transects of $VT_{\mbox{up-seas}}$ (color-filled black contours) for the  Canary (top) and Benguela (bottom) upwelling system, within each sub-region (column). $VT_{\mbox{up-seas}}$ corresponds to an along-shore seasonal average over the entire historical period ($1995-2014$). The three-month seasonal average is adapted to each sub-region to only consider months most favourable to upwelling. Contours of potential temperature (light grey) and mixed layer depth (solid white) are superimposed. Line plots above each transect show cross-shore profiles of Sea Surface Temperature (solid black), offshore Ekman transport (solid red)and $VT_{m}$ (blue).  The model shown here is the HadGEM3-GC3.1-MM model. }
  \label{fig:cell_plot}
\end{figure*}

After validating the seasonal cycle of the indices, we choose the most intense upwelling season for each of the subregions to consider the vertical structure of the upwelling cell in each subregion. The resulting upwelling cells derived from the HadGEM3-GC3.1-MM model can be seen in Fig. \ref{fig:cell_plot}. We display above each cell the Ekman transport, the $VT_{m}$ and the SST. Looking at the upwelling cell, some features are similar to the annual means over the whole region shown in Fig. \ref{fig:cell_annual}. However, the different subregions also show differences in their vertical structures. In the Canary region, maxima are found closer to the coast in the NCUC and the CCUC, while the SCUC shows a higher zonal extent. This corresponds to maximal wind stress intensities close to the coast in the northern and central parts, while in the south maxima are found farther offshore. The highest upwelling rates are found in the SCUC, which has been described to be the most productive region in the Canary Upwelling system \citep{sylla2019weakening}. In the Benguela upwelling the highest upwelling rates and vertical extent can be found in the CBUC, which is where the Lüderitz cell, the most intense upwelling cell in the Benguela system, is located. This is in agreement with \cite{chang2023uncertain}. 
The SBUC shows only slightly lower upwelling rates and vertical extent, even though the transport calculated from wind is much weaker. The maximum in the SBUC is found farther from the coast. This is mirrored in the line plots above the cell plots, where maximum Ekman transport calculated from meridional wind stress and maximum $VT_{m}$ peak at the same distance from the coast. As the Ekman transport resulting from wind stress forcing is the main driver of upwelling, this is an expected outcome, reflecting the ocean response to the atmospheric forcing. The difference between the Ekman transport at the coast and the Ekman transport offshore reflects the importance of the zonal gradient of the meridional wind stress. We have verified that this term is the most important component of the wind stress curl. 

The differences between the subregions in Fig. \ref{fig:cell_plot} can be found in the vertical extent, the distance to the coast of the maximum values and the depth of the maximum values. They are influenced by static features, such as the coastal shape and shelf structure \citep{BENAZZOUZ201438}. This might, for example, be a factor in the maxima occuring farther offshore in the southern regions of both upwellings, as they are characterized by a wider, shallow continental shelf. Dynamical features also play a role, such as the strength and extent of the meridional wind stress forcing, and the Wind Stress Curl (WSC), which have an impact on the amount and depth from which water is upwelled \citep{Capet2004}. For future work, a closer analysis of the coastal bathymetry, coastline shape and WSC generated Ekman pumping on a regional scale in comparison with the divergence of Ekman transport could help explain differences between the depth from which water is upwelled. Regarding the relation between the Ekman Index and the Vertical Transport, the closer the cell is located towards the equator, the larger the difference becomes between the two indices, as already noted in section \ref{seas}. This can have an important implication on upwelling processes as especially geostrophic transport has been found to have a counteracting or enhancing effect on upwelling through convergence or divergence \citep{veitch2010modeling} and could play a large role in controlling the future of upwelling regions \citep{sylla2019weakening, jing2023geostrophic}. \cite{jing2023geostrophic} found especially large differences between their wind-based index and vertical transport based index in the Atlantic Upwelling Systems, which they ascribe to downwelling caused by geostrophic flows. They further state that analysing changes in the geostrophic flow could therefore help in determining future changes. As in Fig. \ref{fig:cell_plot}, the differences between the $EI$ and the $VT_{m}$ are highest in the equatorward portions, suggesting that regional differences of the impact of the ocean circulation within the two Atlantic upwellings could also be an important factor. In the Benguela upwelling, \cite{veitch2010modeling} have found an inhibition of upwelling in the northern part due to geostrophic convergence and an enhancement of upwelling in the southern part due to geostrophic divergence. The geostrophic convergence in the tropical parts of the upwelling regions is ascribed to the eastward equatorial currents of the Angola Dome in the Benguela and the Guinea Dome in the Canary Upwelling system \citep{marchesiello2010upwelling}. This could explain the patterns observed in Fig. \ref{fig:cell_plot}. 

We discuss the main differences between the models briefly (the figures for the remaining four models can be found in the supplementary material, Figs. \ref{fig:S1}-\ref{fig:S4}). The vertical structure of the cells is comparable across models and resolutions. Consistent with higher wind stress as described in section \ref{seas}, in general vertical transport is higher in the HadGEM models (Figs. \ref{fig:cell_plot}, \ref{fig:S1} and \ref{fig:S2}) than the CNRM ones (Figs. \ref{fig:S3} and \ref{fig:S4}) as has been already observed in the seasonal cycle in Figs. \ref{fig:seasonal_cycle_can} and \ref{fig:seasonal_cycle_ben}. 
In the Canary upwelling, the influence of resolution is contrasted between the two models. For the HadGEM model, vertical transport is weaker in all three subregions with higher resolution, while in the CNRM models, transport is also weaker in the NCUC and SCUC in the high resolution model, but stronger in the CCUC. 
In the Benguela, in the HadGEM models the highest intensities in the CBUC and SBUC occur at the ¼ \degree\ resolution. For CNRM, the transport is higher in the low resolution model. 
This behaviour is different from the CESM model used in \cite{chang2023uncertain}, where the lower resolution model shows lower transport in all four major upwelling regions. This indicates that both model and resolution can play an important role in the representation of fine scale differences in upwelling vertical water transport. This has also been investigated in a recent study by \cite{sylla2022} analysing CMIP6 model performance in the Canary upwelling system, which has shown that an efficient method to have an improvement in upwelling representation is increasing both oceanic and atmospheric resolution. However, they also highlight the importance of model parametrization, tuning, spin-up period length and representation of processes.
\newline

\subsection{Linear trends over the 21st century}

In this section, we investigate the behavior of the Canary and Benguela upwelling systems in future projections all run under the  highest emission scenario. 

\begin{figure}[t]
\begin{center}
    \includegraphics[width=0.5\textwidth]{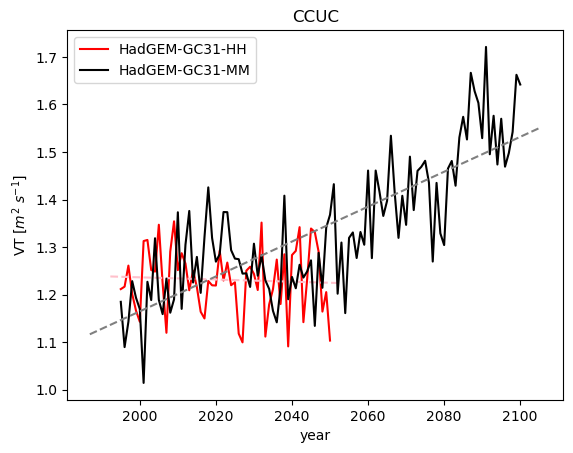}
  \caption{1995-2100 time series of yearly averaged $VTI$ in the CCUC subregion computed from the HadGEM3-GC3.1-HH (solid red) and the HadGEM3-GC3.1-MM (solid black) simulations. The trend lines (dashed) are calculated using linear regression. While HadGEM3-GC3.1-HH shows a trend that is not significant with a p-value of 0.916, the HadGEM3-GC3.1-MM model shows a trend significantly different from 0 with a p-value of 0.006.}
  \label{fig:time_series}
\end{center}
\end{figure}

Fig.~\ref{fig:time_series} shows the 1995-2100 time series of $VTI$ over the CCUC. Although the yearly averaged-values have only been plotted for the HadGEM3-GC3.1-HH and the HadGEM3-GC3.1-MM simulations, the trends illustrate our finding that significant changes are only found during the second half of the 21st century, in agreement with other CMIP6 models \citep{IPCCRep}. Indeed, when calculating the trends for the 1995-2050 period only, none of the models shows significant trends. Additionally, Fig.~\ref{fig:time_series} highlights the importance of the high multi-annual variability which has a strong negative impact on the confidence interval of the computed linear trends.

Trends per century are calculated within each sub-region for the upwelling index $VTI$ presented in section~\ref{sec:indices}. The trends are calculated over the 1995-2100 period from a linear regression performed on the yearly average values of the upwelling indices. The 90\% confidence interval is computed from the student distribution with a number of degrees of freedom chosen after dividing the length of the time series by the corresponding lag associated with the 0.5 crossing of the auto-correlation function ($\simeq 12$ years). Note that the HadGEM3-GC3.1-HH simulation is excluded from our analysis because it only covers the 1995-2050 period. Some tests have been done on the definition of the depth and time scale of the index for the trend calculation. They have shown that for achieving statistically significant results, annual means provide better results than seasonal means. A constant depth is chosen to provide comparability with previous studies. The depth of 50 m having provided results closest to the Ekman index and showing the most significance, this is used for the calculation of the index for the future. 

\begin{figure}[ht]
\begin{center}
  \begin{subfigure}{0.5\textwidth}
    \textbf{(a)}
    \end{subfigure}
  \begin{subfigure}{0.5\textwidth}
    \centering
    \includegraphics[width=\textwidth]{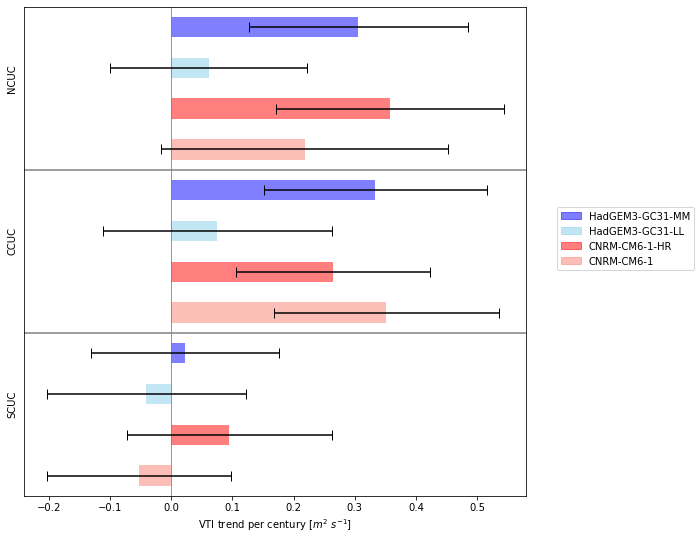}
  \end{subfigure}
    \begin{subfigure}{0.5\textwidth}
    \textbf{(b)}
    \end{subfigure}
  \begin{subfigure}{0.5\textwidth}
    \centering
    \includegraphics[width=\textwidth]{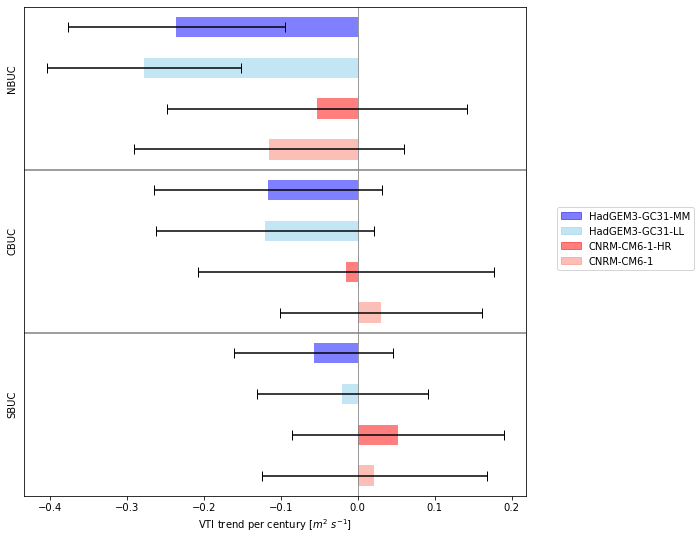}
  \end{subfigure}
  \caption{Trend per century in the Vertical Transport Index calculated over period 1995-2100 from historical simulation and SSP5 – 8.5 runs for each upwelling cell in the (a) Canary and (b) Benguela Upwelling System over annual means for the medium and low resolution models. The error bar shows the 90\% confidence interval.}
  \label{fig:trend}
\end{center}
\end{figure}

Some models show significant trends in vertical transport (Fig. \ref{fig:trend}), although no significant trends have been found in the wind-derived Ekman Index (supplementary material \ref{fig:S10}) and the WSC (not shown here, very similar to trends in Ekman Index). 
Nevertheless, $EI$ and $VTI$ are clearly related at interannual time scales. 
We have computed the Pearson correlation of the entire time series (1995-2050 for the high resolution and 1995-2100 for the medium and low resolution models) for all models and subregions. This has shown very high correlations coefficients (0.56 - 0.97). When comparing the correlation for the historical (1995-2014) and future (2031-2050, 2081-2100) period, changes are very small. This means that wind is and stays one of the main drivers of upwelling. Regarding the subregions, our results show, with only one exception (HadGEM3-GC3.1-LL in the SCUC), that the correlation between the $EI$ and $VTI$ is lower in the equatorward portions of the two Atlantic upwellings (multimodel mean NCUC: 0.93, SCUC: 0.83, NBUC: 0.76, SBUC: 0.96). This confirms the hypothesis that in these regions the complex oceanic circulation plays a  more important role in defining upwelling and its future changes than in the subtropical regions. This is an important factor in understanding why only looking at trends in the wind might provide an incomplete picture. 

The analysis of the computed trends allows us to revisit the two common hypothesis regarding the future of upwelling under global warming. Under the Bakun hypothesis \citep{bakun1990global}, an overall intensification of wind would occur and result in increased upwelling. Under the hypothesis of poleward migration however, we would expect upwelling intensities to increase (decrease) in the poleward (equatorward) subregions \citep{rykaczewski2015poleward, bograd2023climate}. 
Using subregions allows us to assess the regional difference in trends, which is an important factor regarding the poleward migration hypothesis. 
In the Canary system, a significant increase occurs in upwelling transport in the northern and central cell in both 1/4\degree\ resolution models and in the CCUC also for the CNRM-CM6-1 model. In the SCUC, the models show less significant results and conflicting signs between the models. In the Benguela region both HadGEM models show a significant decrease in the northern cell, whereas in the central and southern region, trends differ in sign between models and resolutions and do not pass the significance test. Since in the Benguela region we find some significant decrease in the NBUC, we can reject the Bakun hypothesis. Since not all our results are significant for all the subregions, we can neither accept nor reject the poleward migration hypothesis. Our results, however, support the finding that trends differ regionally. This highlights the sensitivity of the results to the choice of the exact region chosen for the analysis. In two recent works presented by \cite{chang2023uncertain} and \cite{jing2023geostrophic} using the high resolution model Community Earth System Model (CESM) from the CMIP6 project and similar methods for assessing vertical transport, trends of conflicting signs have been found. 

In the Canary region, our results are congruent with the results of \cite{chang2023uncertain} and \cite{jing2023geostrophic}. \cite{chang2023uncertain} use the region from 36\degree N to 15\degree N, while \cite{jing2023geostrophic} look at the region between 32\degree N and 22\degree N. This covers mostly the NCUC and CCUC in our work. Overall positive trends are found in both studies, especially in the central part. This is also the case for our results. The SCUC shows contradicting trends in our work, this region is only partly considered (the northern section) in \cite{chang2023uncertain} and excluded from the region chosen in \cite{jing2023geostrophic}. In the Benguela region, our results suggest a decrease in the NBUC and contradictory trends between the models in the CBUC and SBUC. \cite{jing2023geostrophic} find overall negative trends in upwelling transport in the Benguela region, while \cite{chang2023uncertain} find an increase in upwelling transport. The latitudinal extent of the region is definitely an important factor. \cite{jing2023geostrophic} define their region as the area from 30\degree S to 10\degree S and \cite{chang2023uncertain} from 32\degree S to 22\degree S. However, looking at the latitudinal distribution of the trends shows that even where their study regions overlap in the central Benguela, different results are obtained: positive trend in \cite{chang2023uncertain} and negative trend in \cite{jing2023geostrophic}. This can be explained by the two studies using a different longitudinal extent for their calculation, 50 km offshore for \cite{chang2023uncertain}  and 200 km offshore for \cite{jing2023geostrophic}. The results differ because the vertical transport trends are opposite close to the coast compared with 100 km offshore (Ping Chang, personal communication). This highlights that mapping the vertical transport as a function of depth and distance to the coast, as done in this work, is a key factor when analysing possible future trends as they might differ locally. 
Regarding the relation between the wind-derived and the vertical transport derived index, \cite{jing2023geostrophic} find strong significant trends in their vertical transport index for both the CESM-H ensemble mean and the CMIP6 ensemble mean. For the Ekman index they find lower significant trends in the CESM-H ensemble mean, and very low, non-significant results in the CMIP6 ensemble mean. Our results in Figs. \ref{fig:trend} and \ref{fig:S10} show similar behaviour in significance.

Regarding the lack of significant trends in the main driving mechanism of the upwelling, the wind, and the lack of significance in our $VTI$ compared to the vertical transport index found in \cite{jing2023geostrophic}, two factors can play a role. The method used to calculate statistical significance in our work is very rigorous, which increases robustness, but leads to less of the models passing the significance testing. This is further complicated by the very high interannual variability in intensities and maximum depth and distance. While \cite{jing2023geostrophic} use ensemble means, we investigate the evolution of the vertical transport of single model members. But, as these processes are subject to high interannual variability in global coupled models, we suggest that to see significant trends, more members or models, alongside long time series until at least the end of the century are needed. 
\newline 

\section{Conclusions and perspectives}
With this work we aim, in a multi-model perspective including eddying ocean components, to address spatial and seasonal variations of upwelling and their future trends in the Canary and Benguela upwelling systems. Further, we explore the use of the vertical water mass transport to map the cross-shore and vertical structure of the upwelling cells. For this we use data from the CMIP6 HadGEM3-GC3.1 and CNRM-CM6-1 models with different spatial resolutions and evaluate them against CCI SST and ERA5 wind stress data. We computed two classical upwelling indices, the SST index and Ekman index as well as an index derived from the integrated vertical water mass transport, the Vertical transport index. We have evaluated the seasonal cycle from those indices in the two Atlantic upwellings. Our analysis shows an accurate representation of the seasonal cycle within each of the subregions for the Ekman and Vertical Transport Index. The SST index mostly captures the onset of upwelling, but the magnitude is largely underestimated. We have focused on the vertical transport, which is the most direct measure of upwelling, and even though vertical processes in models are subject to some noise \citep{IPCCRep}, integrating the transport and looking at subregions provides a clear and consistent picture of the upwelling cell.
Our results highlight differences between the subregions regarding maximum upwelling depth, distance to the coast and vertical extent. Especially the difference in maximum depth and vertical extent can influence the origin of upwelled water, which has a direct impact on the ecosystem \citep{IPCCRep}. Studying the respective impact of wind stress and WSC on a regional scale could help understand these differences better. For this a regional model forced with CMIP6 output could be a next step. This approach has previously been used to assess local impacts of climate change \citep{chamorro2021projection} and local wind-structure \citep{small2015}.  
When looking at future trends in the vertical transport, we find a significant increase in transport in the northern and central Canary region in both medium resolution models and the CNRM low resolution model, while the southern Canary region shows contradicting trends. In the Benguela region we find some significant negative trends in the northern region, while the central and southern part differ between models. While we find some significant trends in the VT, we do not find significant trends in the wind derived indices. This means that other processes likely play a role in future changes. Our work further highlights that the definition of the subregions or study region can have an important impact on the results of future trends. \cite{chang2023uncertain} and \cite{jing2023geostrophic} obtained different signs in trends using the same model and comparable analysis methods for the Benguela region, probably due to different spatial extents in latitude and longitude. \cite{jing2023geostrophic} have linked future changes to geostrophic flow. The importance of these processes does however also highlight the advantages of an ocean-derived index as is the vertical transport, since it reflects all important dynamics and their changes by directly showing the ocean response to external forcing, therefore also reflecting the wind forcing, which will stay one of the major impacting factors in the Atlantic EBUS future. Once significant trends can be established from vertical transport, it can then become a very powerful tool in assessing and understanding future changes. This can help answer some of the key questions regarding the future of EBUS, including the impact of geostrophic and ageostrophic transport \citep{jing2023geostrophic} and stratification \citep{bograd2023climate}. As the new phase of CMIP goes into planning, the intended output of wmo as a baseline variable will be an important asset in the availability of this variable \citep{juckes}. Together with improvements in its representation, this could provide new pathways in analysing EBUS.

In this work we focus on the behaviour of each model and the possibility of using the vertical water mass transport for projections, rather than providing a complete picture of the future of upwelling systems. Even though some significant trends can be seen even when only looking at one model realization, for a more in-depth analysis the use of more ensemble members and more models could give a better insight and more reliable results especially for ambiguous regions like the southern Benguela. More ensemble members could further help increasing statistical significance of trends of the upwelling favourable winds. 

We observe an improvement in performance with higher resolution. Therefore, particularly the well-performing high resolution simulations should be extended to provide projections until the end of the 21st century, as this is when most changes are expected to be observed in upwelling regions. This is planned in the next phase of HighResMIP and in the European Eddy-Rich Earth System Models project. The availability of multiple eddy-resolving models with more ensemble members will reduce uncertainties and allow more reliable future projections of upwelling systems.








\section{Acknowledgments} \label{acknow}
CMIP6: We acknowledge the World Climate Research Programme, which, through its Working Group on Coupled Modelling, coordinated and promoted CMIP6. We thank the climate modeling groups for producing and making available their model output, the Earth System Grid Federation (ESGF) for archiving the data and providing access, and the multiple funding agencies who support CMIP6 and ESGF.  This work is supported by the MixED Layer hEterogeneitY (MEDLEY) project funded by JPI Climate and JPI Oceans under the 2019 Joint Call, managed by the French Agence Nationale de la Recherche, under contract 19-JPOC-0001-01. It is also a contribution to the EERIE project (Eddy Rich Earth System Models) funded by the European Union (Grant Agreement No 101081383). Raquel Flügel received a Ph.D. grant from the University of Brest and Region Bretagne in the context of ISblue (ANR-17-EURE-0015).

\clearpage


\clearpage

\clearpage

\clearpage

\clearpage

\clearpage

\clearpage

\clearpage

\clearpage

\begin{table*}[h] 
\caption{CMIP6 models used in this work, their horizontal resolution and number of vertical layers used for the atmospheric and oceanic component \citep{IPCC2021annex}.}\label{tab:models}
\begin{tabular*}{\tblwidth}{@{}CCCCC@{}}
\toprule
 Model & Average Atmosphere Resolution & Average Ocean Resolution & Dataset reference \\  \\  
\midrule

        HadGEM3-GC31-LL & 140km, 85L & 70km, 75L & \cite[{[dataset]}][]{Jeff2018}\\ 
        HadGEM3-GC31-MM & 60km, 85L & 20km, 75L & \cite[{[dataset]}][]{Ridley2019}\\ 
        HadGEM3-GC31-HH & 30km, 85L & 7km, 75L & \cite[{[dataset]}][]{Roberts2018}\\ 
        CNRM-CM6-1 &  140km, 91L & 70km, 75L & \cite[{[dataset]}][]{Voldoire2019}\\
        CNRM-CM6-1-HR & 50km, 91L &  20km, 75L & \cite[{[dataset]}][]{Voldoire2019b}\\
\bottomrule
\end{tabular*}
\end{table*}

\clearpage
\bibliographystyle{cas-model2-names}

\bibliography{cas-refs}



\newpage
\section*{Supplementary material}
\renewcommand\thefigure{S.\arabic{figure}}
\setcounter{figure}{0} 
\begin{figure*}[ht]
    \includegraphics[width=\textwidth]{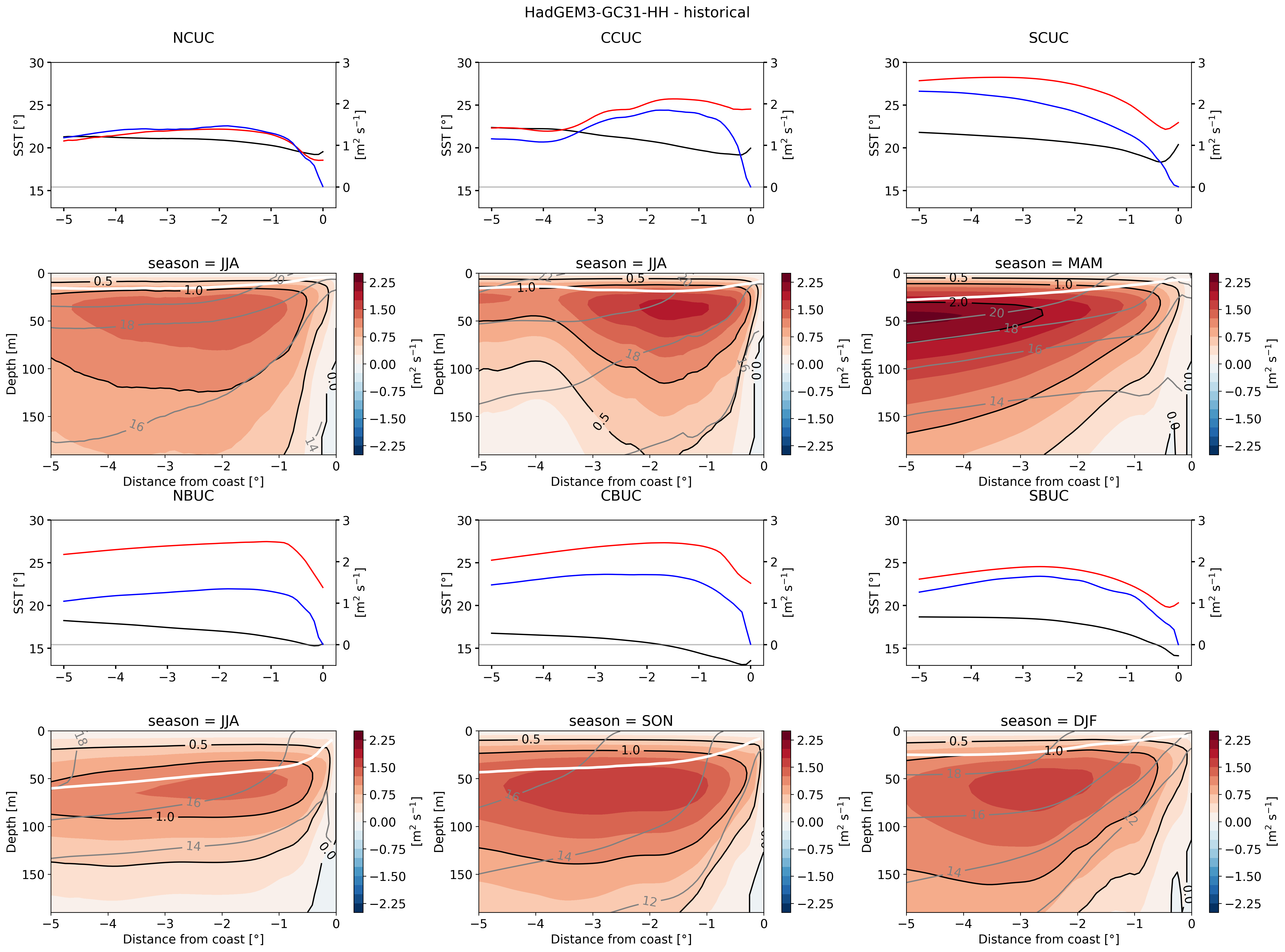}
  \caption{ Cross-shore transects of $VT_{\mbox{up-seas}}$ (color-filled black contours) for the  Canary (top) and Benguela (bottom) upwelling system, within each sub-region (column). $VT_{\mbox{up-seas}}$ corresponds to an along-shore seasonal average over the entire historical period ($1995-2014$). The three-month seasonal average is adapted to each sub-region to only consider months most favourable to upwelling. Contours of potential temperature (light grey) and mixed layer depth (solid white) are superimposed. Line plots above each transect show cross-shore profiles of Sea Surface Temperature (solid black), offshore Ekman transport (solid red)and $VT_{m}$ (blue). The model shown here is the HadGEM3-GC3.1-HH model.}
  \label{fig:S1}
\end{figure*}

\begin{figure*}[ht]
    \includegraphics[width=\textwidth]{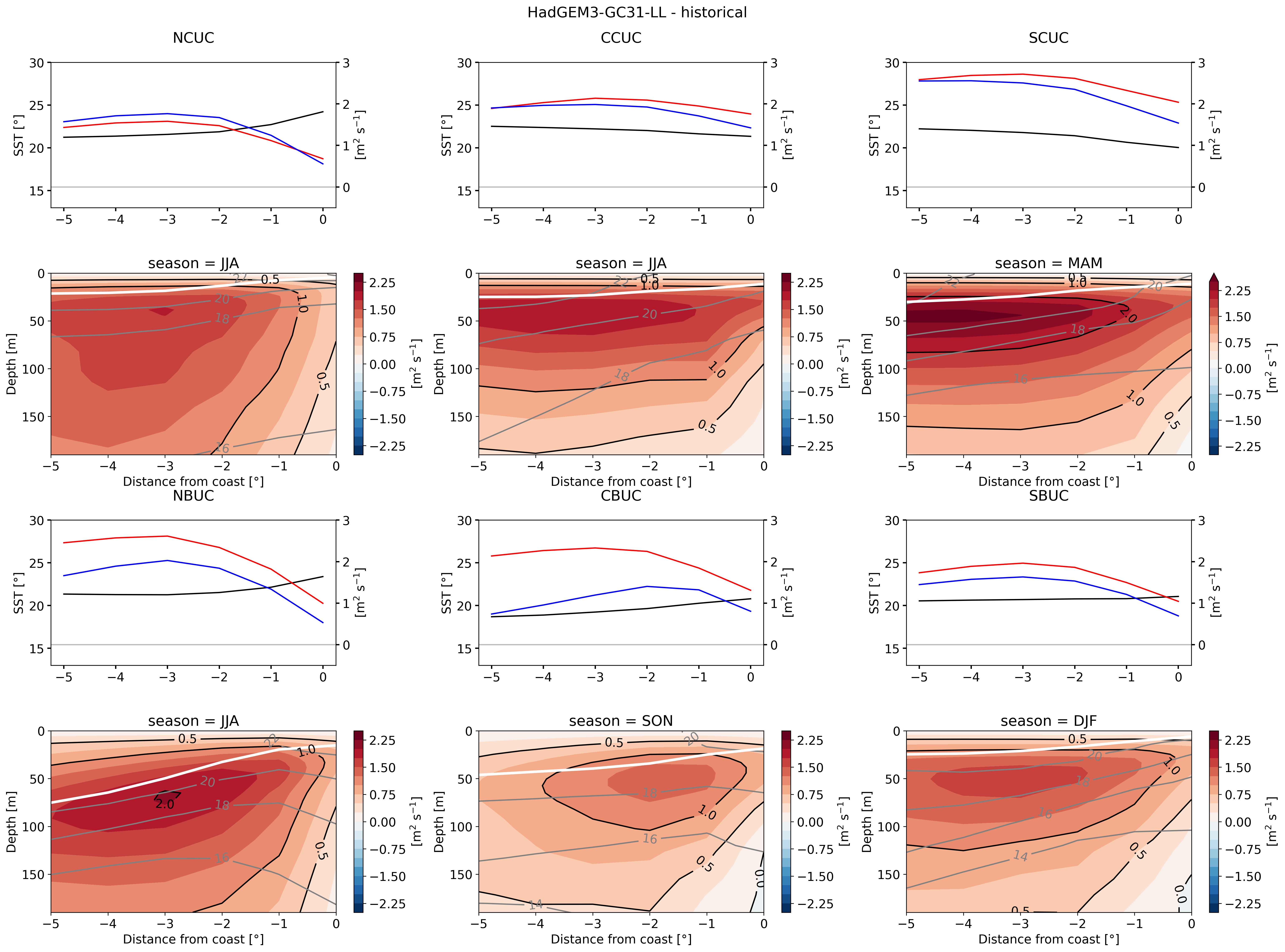}
  \caption{ Cross-shore transects of $VT_{\mbox{up-seas}}$ (color-filled black contours) for the  Canary (top) and Benguela (bottom) upwelling system, within each sub-region (column). $VT_{\mbox{up-seas}}$ corresponds to an along-shore seasonal average over the entire historical period ($1995-2014$). The three-month seasonal average is adapted to each sub-region to only consider months most favourable to upwelling. Contours of potential temperature (light grey) and mixed layer depth (solid white) are superimposed. Line plots above each transect show cross-shore profiles of Sea Surface Temperature (solid black), offshore Ekman transport (solid red)and $VT_{m}$ (blue). The model shown here is the HadGEM3-GC3.1-LL model.}
    \label{fig:S2}
\end{figure*}

\begin{figure*}[ht]
    \includegraphics[width=\textwidth]{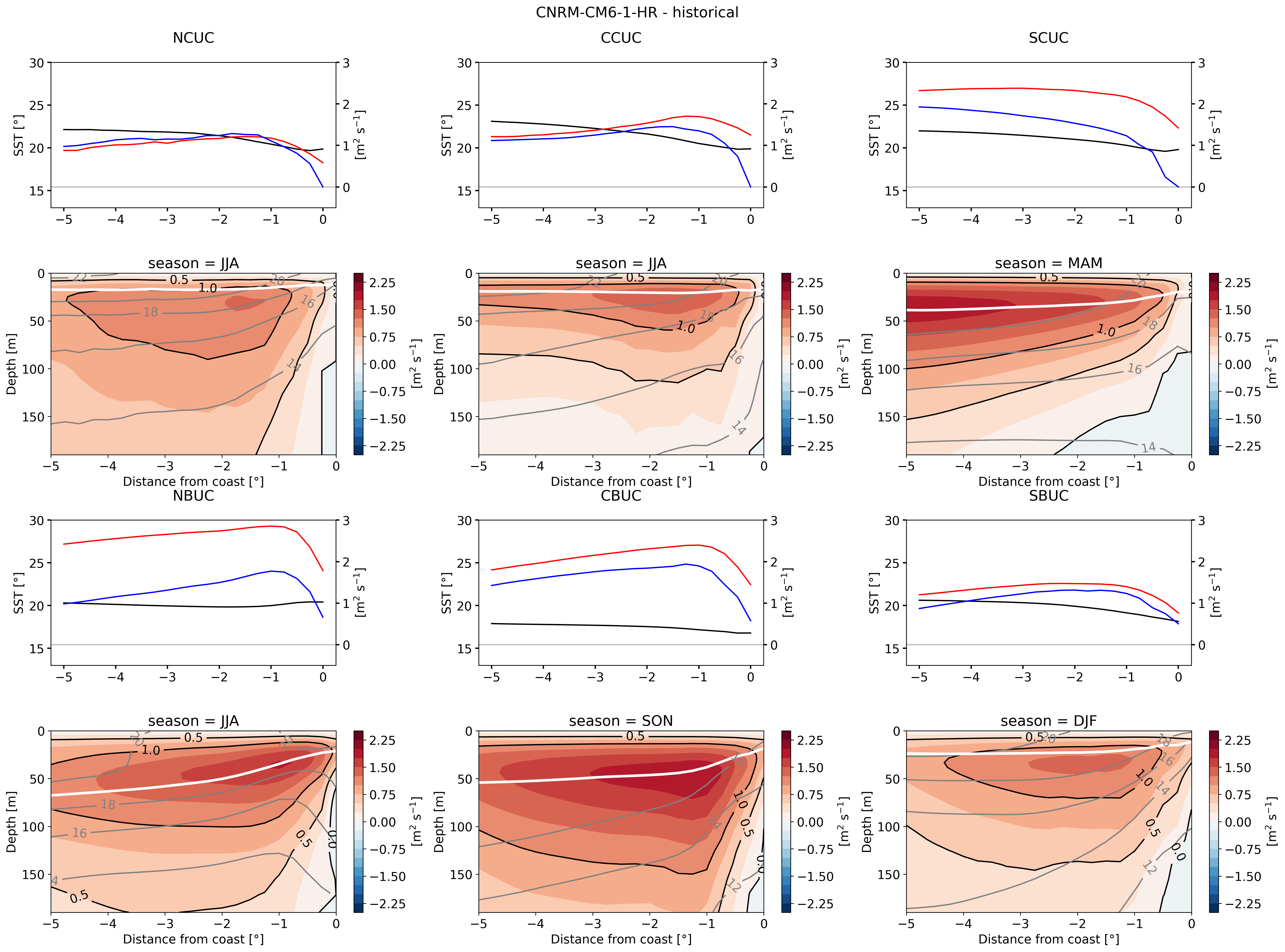}
  \caption{ Cross-shore transects of $VT_{\mbox{up-seas}}$ (color-filled black contours) for the  Canary (top) and Benguela (bottom) upwelling system, within each sub-region (column). $VT_{\mbox{up-seas}}$ corresponds to an along-shore seasonal average over the entire historical period ($1995-2014$). The three-month seasonal average is adapted to each sub-region to only consider months most favourable to upwelling. Contours of potential temperature (light grey) and mixed layer depth (solid white) are superimposed. Line plots above each transect show cross-shore profiles of Sea Surface Temperature (solid black), offshore Ekman transport (solid red)and $VT_{m}$ (blue). The model shown here is the CNRM-CM6-1-HR model.}
    \label{fig:S3}
\end{figure*}

\begin{figure*}[ht]
    \includegraphics[width=\textwidth]{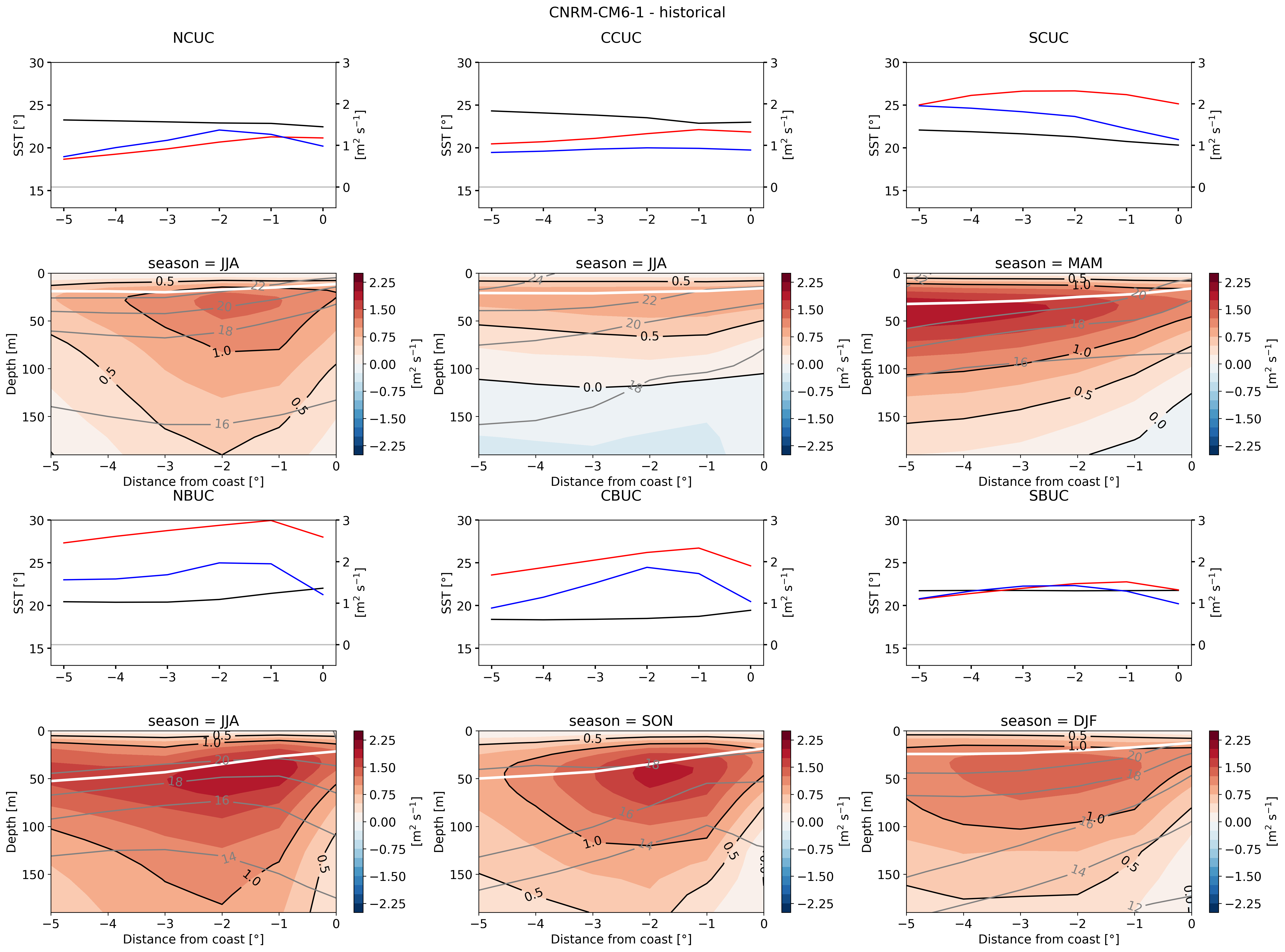}
  \caption{ Cross-shore transects of $VT_{\mbox{up-seas}}$ (color-filled black contours) for the  Canary (top) and Benguela (bottom) upwelling system, within each sub-region (column). $VT_{\mbox{up-seas}}$ corresponds to an along-shore seasonal average over the entire historical period ($1995-2014$). The three-month seasonal average is adapted to each sub-region to only consider months most favourable to upwelling. Contours of potential temperature (light grey) and mixed layer depth (solid white) are superimposed. Line plots above each transect show cross-shore profiles of Sea Surface Temperature (solid black), offshore Ekman transport (solid red)and $VT_{m}$ (blue). The model shown here is the CNRM-CM6-1 model.}
    \label{fig:S4}
\end{figure*}

\begin{figure*}[ht]
    \includegraphics[width=\textwidth]{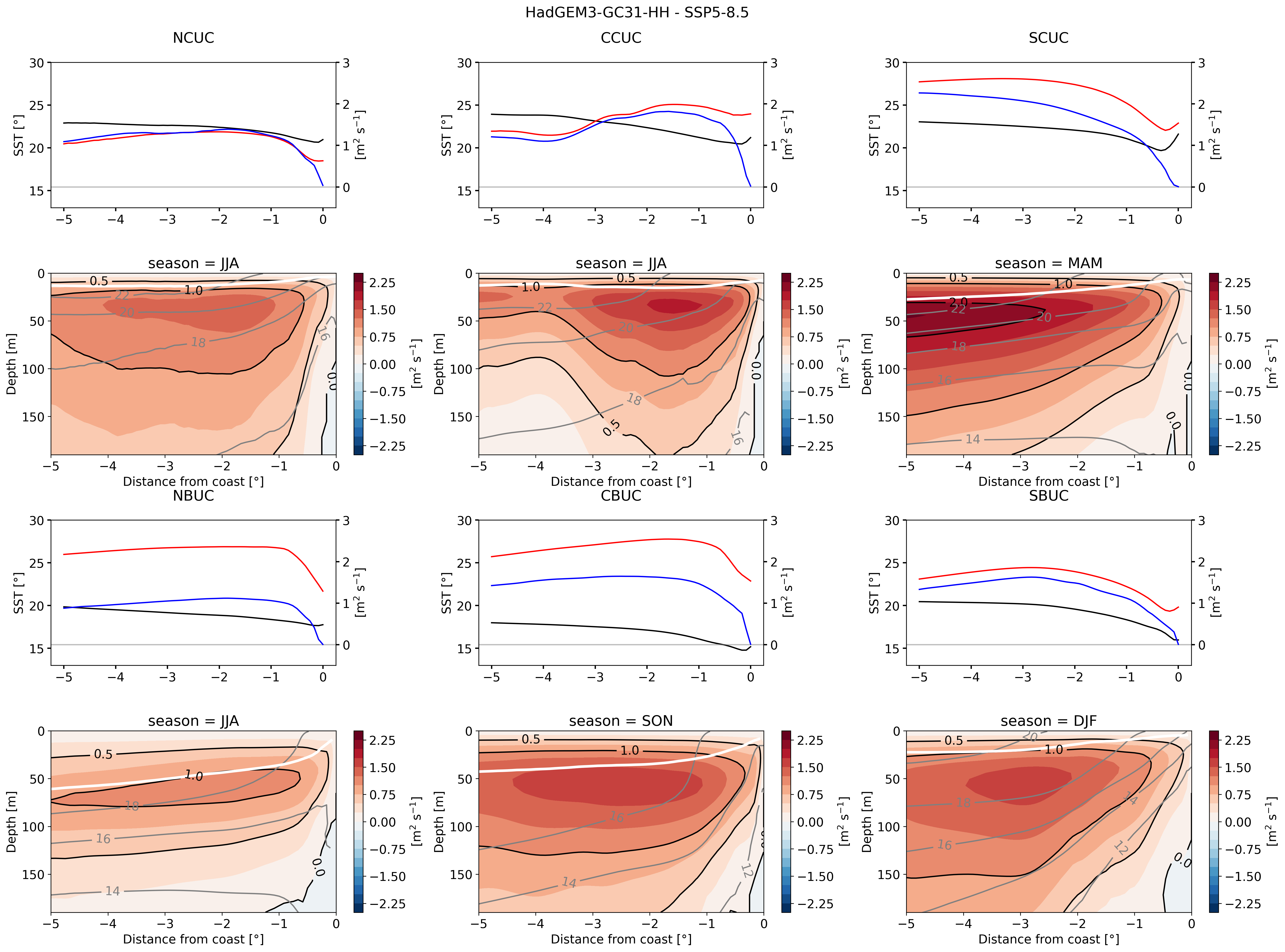}
  \caption{ Cross-shore transects of $VT_{\mbox{up-seas}}$ (color-filled black contours) for the  Canary (top) and Benguela (bottom) upwelling system, within each sub-region (column). $VT_{\mbox{up-seas}}$ corresponds to an along-shore seasonal average over the entire future period ($2031-2050$). The three-month seasonal average is adapted to each sub-region to only consider months most favourable to upwelling. Contours of potential temperature (light grey) and mixed layer depth (solid white) are superimposed. Line plots above each transect show cross-shore profiles of Sea Surface Temperature (solid black), offshore Ekman transport (solid red)and $VT_{m}$ (blue). The model shown here is the HadGEM3-GC3.1-HH model.}
    \label{fig:S5}
\end{figure*}

\begin{figure*}[ht]
    \includegraphics[width=\textwidth]{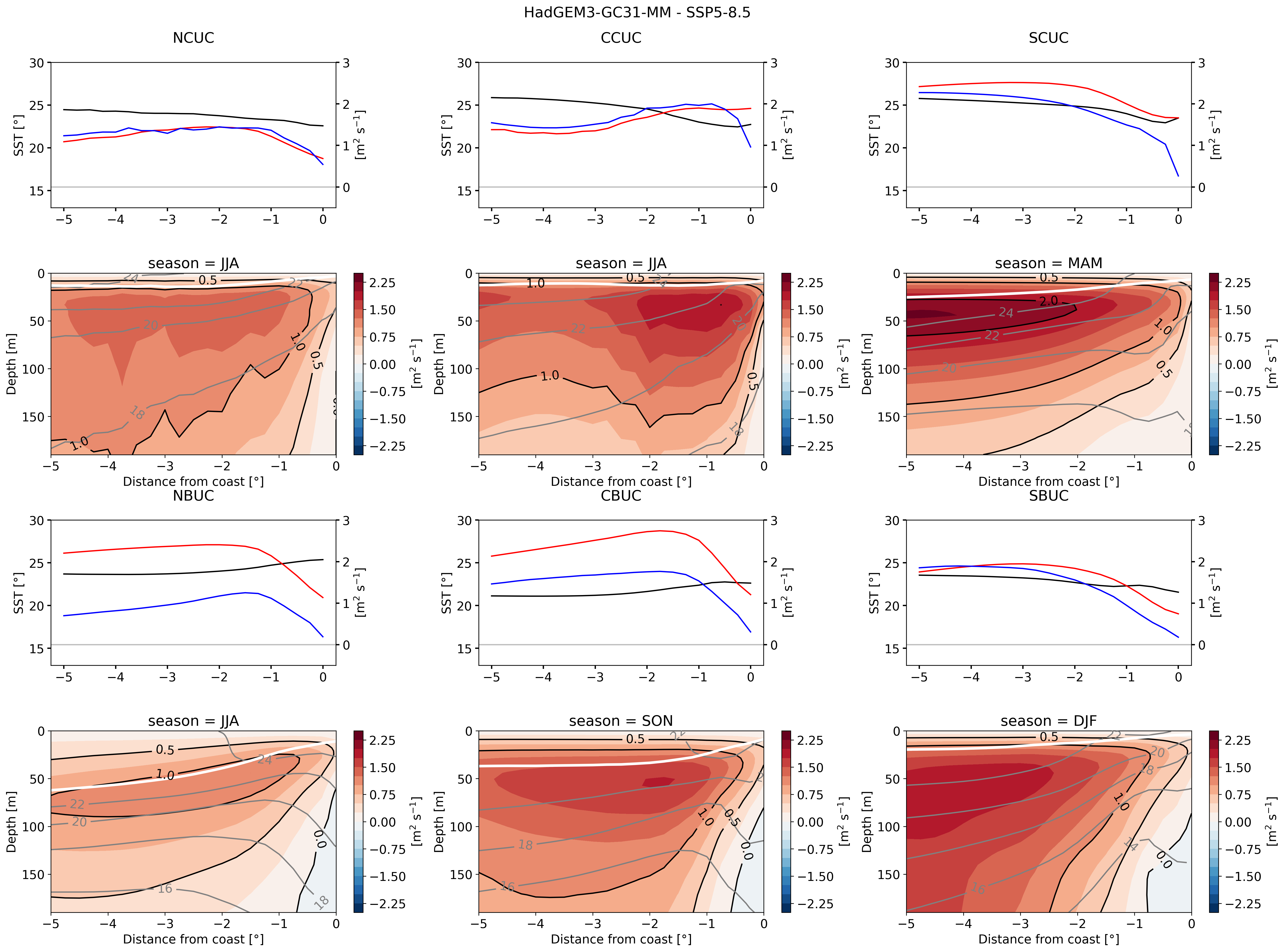}
  \caption{ Cross-shore transects of $VT_{\mbox{up-seas}}$ (color-filled black contours) for the  Canary (top) and Benguela (bottom) upwelling system, within each sub-region (column). $VT_{\mbox{up-seas}}$ corresponds to an along-shore seasonal average over the entire future period ($2081-2100$). The three-month seasonal average is adapted to each sub-region to only consider months most favourable to upwelling. Contours of potential temperature (light grey) and mixed layer depth (solid white) are superimposed. Line plots above each transect show cross-shore profiles of Sea Surface Temperature (solid black), offshore Ekman transport (solid red)and $VT_{m}$ (blue). The model shown here is the HadGEM3-GC3.1-MM model.}
    \label{fig:S6}
\end{figure*}

\begin{figure*}[ht]
    \includegraphics[width=\textwidth]{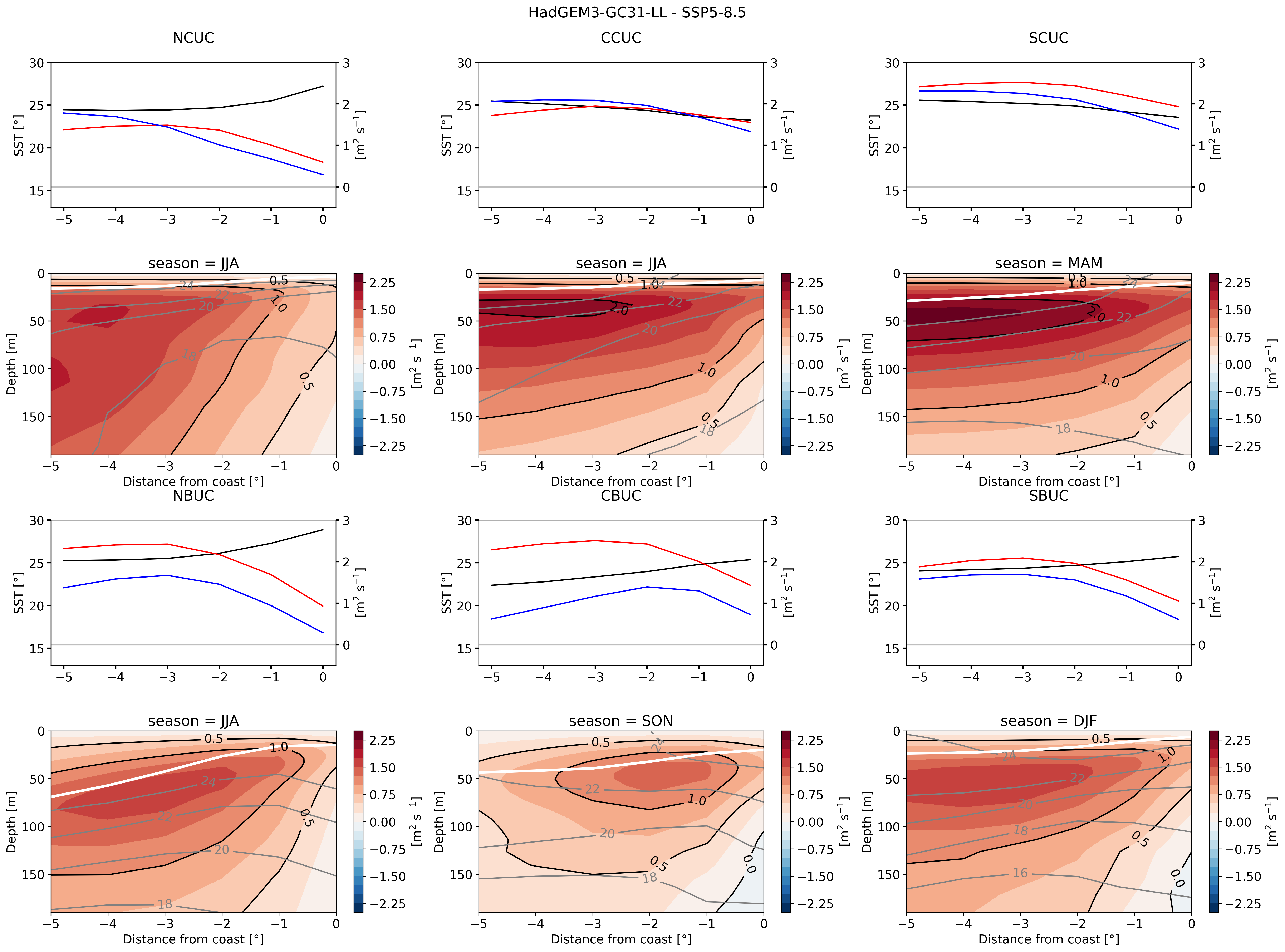}
  \caption{Cross-shore transects of $VT_{\mbox{up-seas}}$ (color-filled black contours) for the  Canary (top) and Benguela (bottom) upwelling system, within each sub-region (column). $VT_{\mbox{up-seas}}$ corresponds to an along-shore seasonal average over the entire future period ($2081-2100$). The three-month seasonal average is adapted to each sub-region to only consider months most favourable to upwelling. Contours of potential temperature (light grey) and mixed layer depth (solid white) are superimposed. Line plots above each transect show cross-shore profiles of Sea Surface Temperature (solid black), offshore Ekman transport (solid red)and $VT_{m}$ (blue). The model shown here is the HadGEM3-GC3.1-LL model.}
    \label{fig:S7}
\end{figure*}

\begin{figure*}[ht]
    \includegraphics[width=\textwidth]{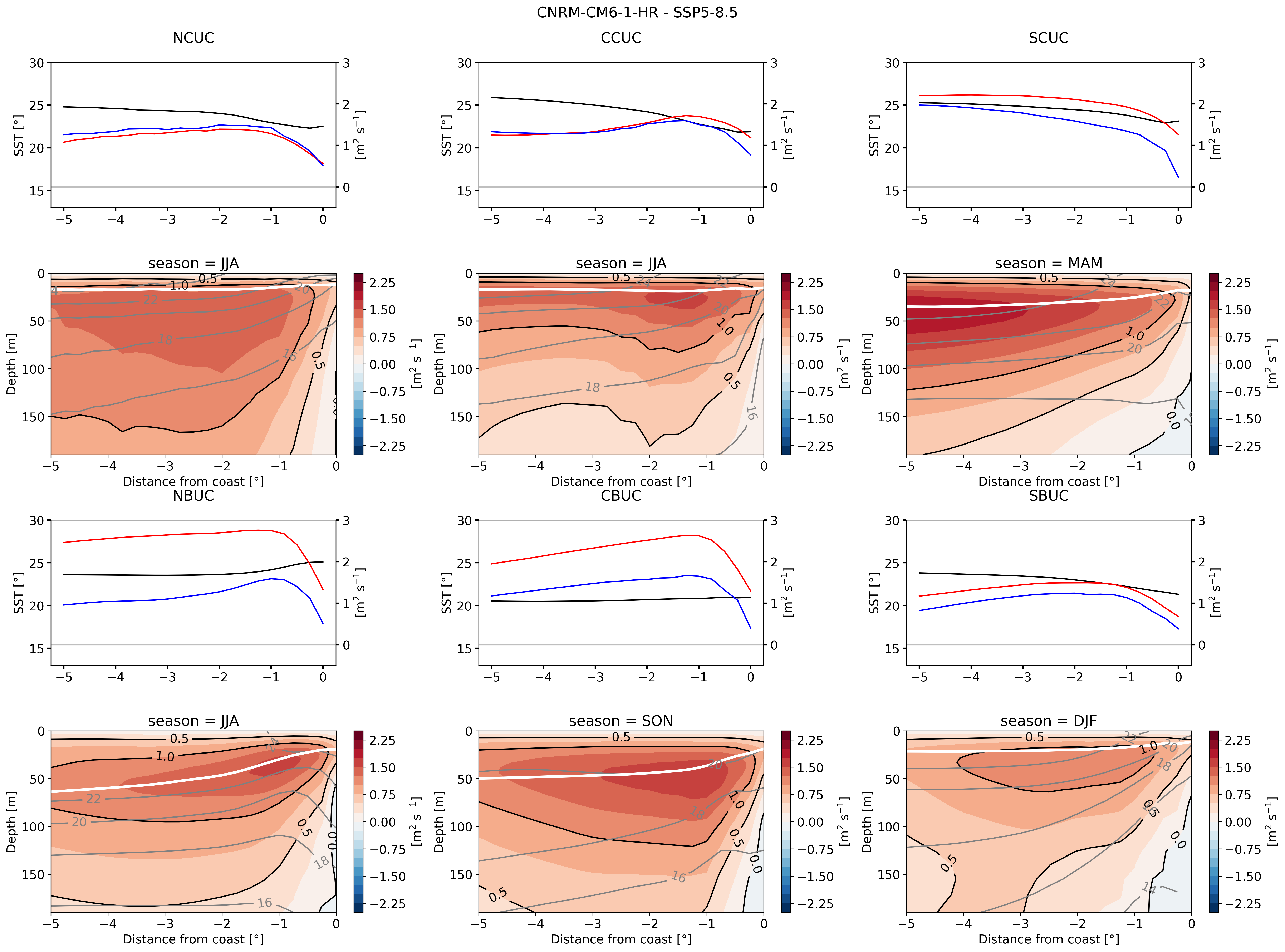}
  \caption{ Cross-shore transects of $VT_{\mbox{up-seas}}$ (color-filled black contours) for the  Canary (top) and Benguela (bottom) upwelling system, within each sub-region (column). $VT_{\mbox{up-seas}}$ corresponds to an along-shore seasonal average over the entire future period ($2081-2100$). The three-month seasonal average is adapted to each sub-region to only consider months most favourable to upwelling. Contours of potential temperature (light grey) and mixed layer depth (solid white) are superimposed. Line plots above each transect show cross-shore profiles of Sea Surface Temperature (solid black), offshore Ekman transport (solid red)and $VT_{m}$ (blue). The model shown here is the CNRM-CM6-1-HR model.}
    \label{fig:S8}
\end{figure*}

\begin{figure*}[ht]
    \includegraphics[width=\textwidth]{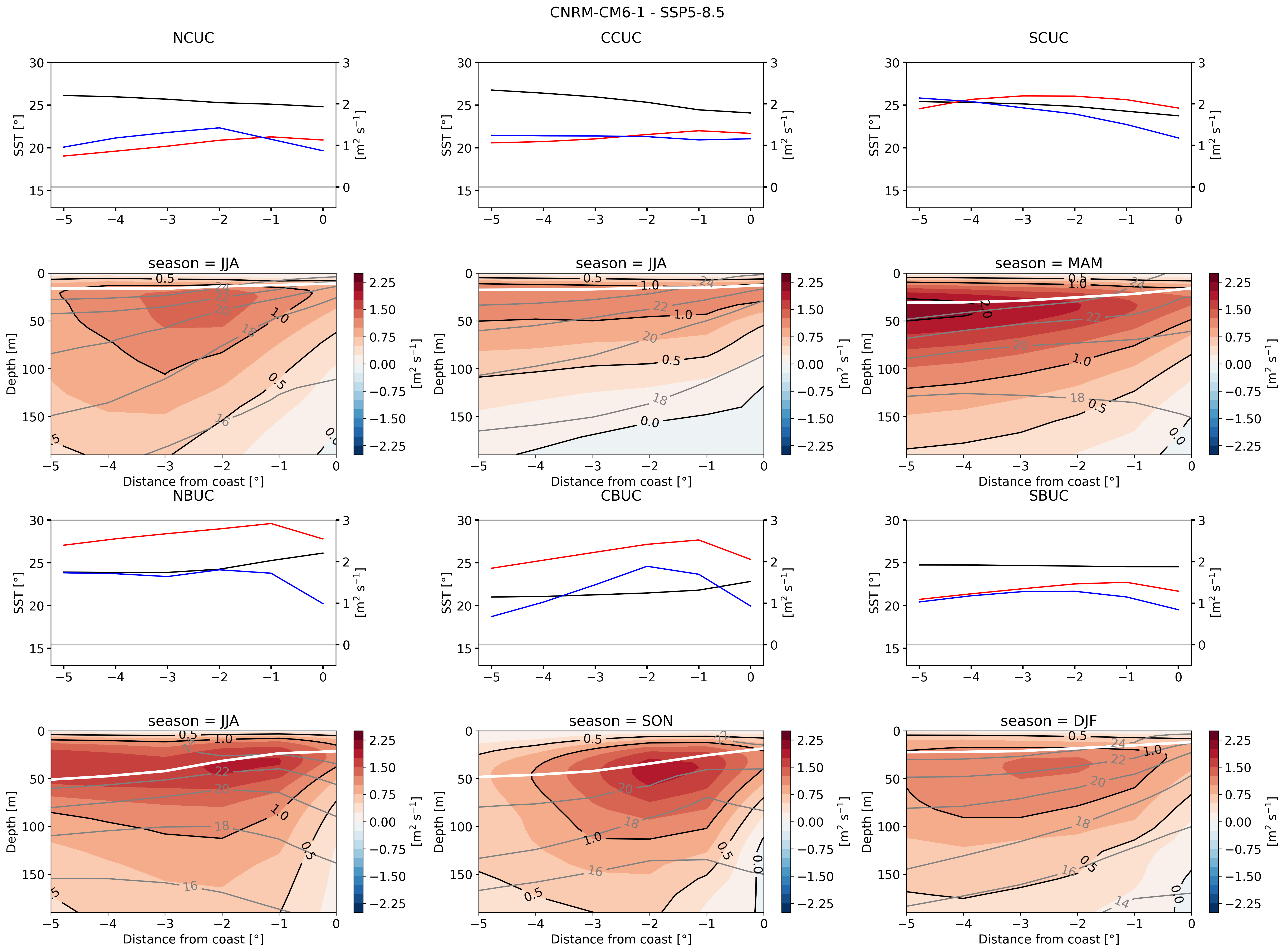}
  \caption{ Cross-shore transects of $VT_{\mbox{up-seas}}$ (color-filled black contours) for the  Canary (top) and Benguela (bottom) upwelling system, within each sub-region (column). $VT_{\mbox{up-seas}}$ corresponds to an along-shore seasonal average over the entire future period ($2081-2100$). The three-month seasonal average is adapted to each sub-region to only consider months most favourable to upwelling. Contours of potential temperature (light grey) and mixed layer depth (solid white) are superimposed. Line plots above each transect show cross-shore profiles of Sea Surface Temperature (solid black), offshore Ekman transport (solid red)and $VT_{m}$ (blue). The model shown here is the CNRM-CM6-1 model.}
    \label{fig:S9}
\end{figure*}

\begin{figure}[ht]
\begin{center}
  \begin{subfigure}{0.5\textwidth}
    \textbf{(a)}
    \end{subfigure}
  \begin{subfigure}{0.5\textwidth}
    \centering
    \includegraphics[width=\textwidth]{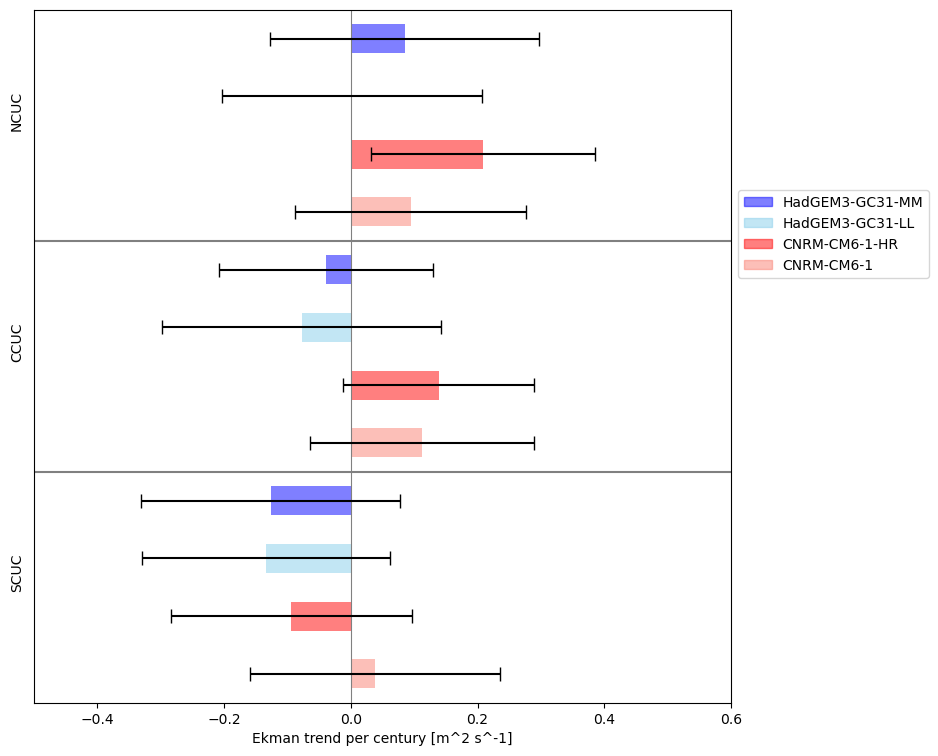}
  \end{subfigure}
    \begin{subfigure}{0.5\textwidth}
    \textbf{(b)}
    \end{subfigure}
  \begin{subfigure}{0.5\textwidth}
    \centering
    \includegraphics[width=\textwidth]{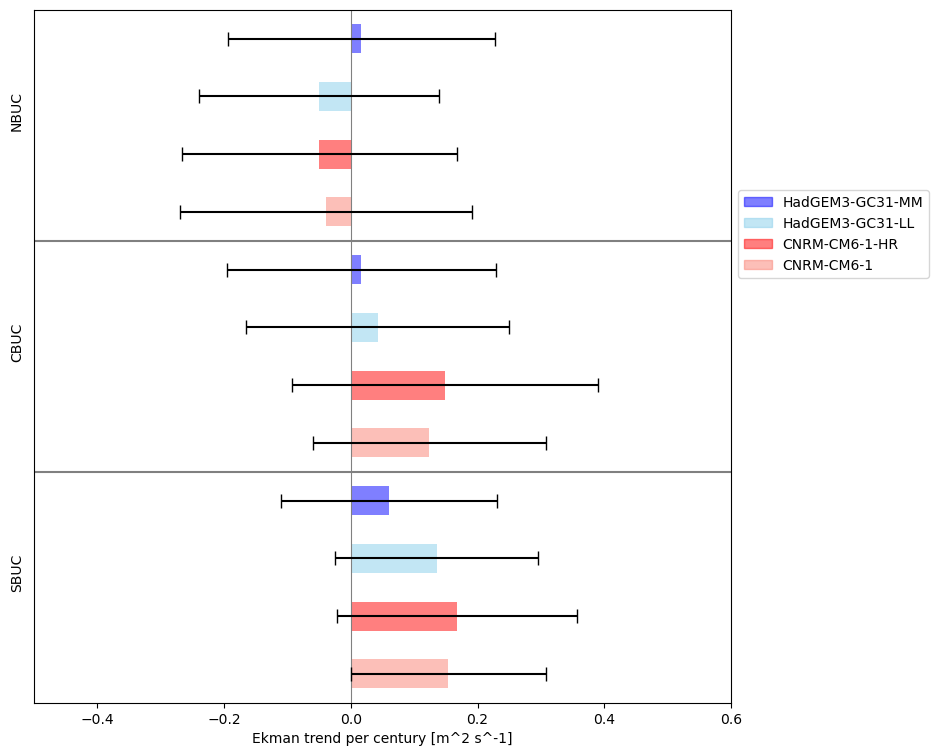}
  \end{subfigure}
  \caption{Trend per century in Ekman transport derived from meridional wind stress calculated over period 1995-2100 from historical simulation and SSP5 – 8.5 runs for each upwelling cell in the (a) Canary and (b) Benguela Upwelling region over annual means for the medium and low resolution models. The error bar shows the 90\% confidence delimitation.}
    \label{fig:S10}
\end{center}
\end{figure}

\end{document}